\documentclass[lettersize,journal]{IEEEtran}
\usepackage{array}
\usepackage{textcomp}
\usepackage{stfloats}
\usepackage{url}
\usepackage{verbatim}
\usepackage{graphicx}
\usepackage{cite}
\usepackage{amsmath,amssymb,amsfonts}
\usepackage{amsthm}
\usepackage{textcomp}  
\usepackage{mathtools} % Bonus
\usepackage{diagbox}
\usepackage{hyperref}
\usepackage{amsmath}
\usepackage{algpseudocode}
\usepackage{caption} 
\usepackage{xcolor}
\usepackage{setspace}
\usepackage{bm}
\usepackage{enumerate}
\usepackage{algorithm}
\usepackage{algpseudocode}

\usepackage{multirow}
\usepackage{subfigure}
\usepackage{bm} 
\usepackage{makecell}

\usepackage{multirow}
\begin{document}

\title{Device Scheduling and Assignment \\[-0pt] in Hierarchical Federated Learning for\\[-0pt] Internet of Things}

% \title{Hierarchical Federated Learning with clustering-based device scheduling and DRL-based device assignment}

\author{Tinghao Zhang, Kwok-Yan Lam, \emph{Senior Member}, IEEE, Jun Zhao\vspace{-10pt}
\thanks{T. Zhang, K. Lam, J. Zhao are with the Strategic Centre for Research in Privacy-Preserving Technologies and Systems (SCRIPTS), and the Digital Trust Centre (DTC), Singapore. The authors are also with the School of Computer Science and Engineering at Nanyang Technological University, Singapore. (Emails: tinghao001@e.ntu.edu.sg; kwokyan.lam@ntu.edu.sg; junzhao@ntu.edu.sg).

This research is supported by the National Research Foundation, Singapore and Infocomm Media Development Authority under its Trust Tech Funding Initiative and Strategic Capability Research Centres Funding Initiative. Any opinions, findings and conclusions or recommendations expressed in this material are those of the author(s) and do not reflect the views of National Research Foundation, Singapore and Infocomm Media Development Authority.

Copyright (c) 20xx IEEE. Personal use of this material is permitted. However, permission to use this material for any other purposes must be obtained from the IEEE by sending a request to pubs-permissions@ieee.org.
}
}

\maketitle
\begin{abstract}
Federated Learning (FL) is a promising machine learning approach for Internet of Things (IoT), but it has to address network congestion problems when the population of IoT devices grows. Hierarchical FL (HFL) alleviates this issue by distributing model aggregation to multiple edge servers. Nevertheless, the challenge of communication overhead remains, especially in scenarios where all IoT devices simultaneously join the training process. For scalability, practical HFL schemes select a subset of IoT devices to participate in the training, hence the notion of device scheduling. In this setting, only selected IoT devices are scheduled to participate in the global training, with each of them being assigned to one edge server. Existing HFL assignment methods are primarily based on search mechanisms, which suffer from high latency in finding the optimal assignment. This paper proposes an improved K-Center algorithm for device scheduling and introduces a deep reinforcement learning-based approach for assigning IoT devices to edge servers. Experiments show that scheduling 50\% of IoT devices is generally adequate for achieving convergence in HFL with much lower time delay and energy consumption. In cases where reduction in energy consumption (such as in Green AI) and reduction of messages (to avoid burst traffic) are key objectives, scheduling 30\% IoT devices allows a substantial reduction in energy and messages with similar model accuracy.
\vspace{-0em}
\end{abstract}

\begin{IEEEkeywords}
Digital trust, privacy-preserving techniques, Hierarchical federated learning, deep reinforcement learning
\end{IEEEkeywords}

\maketitle

\section{Introduction}

The advent of 5G networks and the Internet of Things (IoT) has enabled connected devices to perform complex tasks empowered by machine learning (ML) techniques. Typically deployed in the forms of Cyber-Physical Systems with sensing and actuating capabilities, the control and operations of such connected devices may be automated by artificial intelligence (AI) based on machine-learned models~\cite{Xuemin22, Wu22}. To train these ML models, IoT devices collect the required data and transmit them to a remote server for centralized training.

However, for applications involving private and confidential sensing data, the transmission of device data raises privacy concerns~\cite{McMahan17,Liu23,WANG21,Huang22}. Besides, centralized training will result in serious computation bottlenecks and prohibitive communication overheads, especially when dealing with a massive amount of training data collected from a large number of IoT devices~\cite{Wen23}. To surmount these obstacles, Federated Learning (FL) has been proposed as a promising solution for training ML models in a distributed manner~\cite{McMahan17}. By leveraging FL, IoT devices can train the model with their data not leaving the devices. After local training on the device, model parameters are uploaded for aggregation. Since the training data are retained on the IoT devices, FL mitigates the risk of a privacy breach at the central server. Thanks to these benefits, modern IoT systems have extensively adopted FL in a range of emerging applications such as Smart Healthcare~\cite{Elayan22}, Smart Agriculture~\cite{Othmane22}, and Smart City~\cite{Qolomany20}.

Nevertheless, the scalability of bandwidth resources of a single remote aggregation server remains a non-trivial issue when accommodating a large number of IoT devices in the training process, hence resulting in network congestion. To overcome this challenge, a new FL approach called Hierarchical Federated Learning (HFL) has been introduced~\cite{Liu20}. In HFL, which typically involves multiple edge servers and a cloud server, IoT devices train their models locally and send model parameters to their assigned edge servers for edge aggregation. After multiple edge updates, the edge servers transmit the edge models to the cloud server for global aggregation. If designed and implemented carefully, HFL can achieve a remarkable reduction in energy consumption and much lower time delay in network communications. In cases where reduction in energy consumption (such as in Green AI~\cite{su13168952}) and reduction of messages in each round (so as to avoid burst traffic) are key objectives, HFL is an attractive direction to explore and deserves in-depth investigation.

Although HFL offers convincing benefits, the deployment of HFL still faces practical challenges. For example, HFL still causes notable communication and computation overheads when all the IoT devices participate in the training process. In addition, the straggler effect will significantly increase the latency of HFL. Specifically, suppose one of the edge servers experiences high latency in finishing its edge aggregation due to a straggler IoT device. In that case, the overall HFL training speed will be negatively affected as the cloud server performs global aggregation only when all edge servers have completed their respective edge aggregations.

In this case, device scheduling is an effective approach to speed up HFL training. Device scheduling refers to the process of selecting a subset of devices to participate in each global iteration. Device scheduling has the potential to reduce energy consumption for training HFL~\cite{Zeng20} and decrease the probability of having a straggler IoT device~\cite{Shi20}. However, non-identically independently distributed (non-IID) datasets pose another challenge in ensuring model accuracy. If not handled properly, the local models trained on the non-IID dataset will become biased towards a certain class, resulting in poorer performance on the other classes. Such biased performance may be propagated to the global model through model aggregation, resulting in slower growth of testing accuracy and requiring more global iterations to converge to a preset target accuracy. Moreover, existing scheduling methods focus on traditional FL and can hardly be expanded into HFL. Therefore, an effective device scheduling algorithm is necessary for HFL to minimize the system cost while warranting learning performance.

Another issue of HFL is device assignment, which refers to the task of assigning the scheduled IoT devices to the edge servers. Device assignment aims to balance the workload across edge servers and ensure that each edge server is not overwhelmed with too many IoT devices to handle. To address this issue, an iterative searching algorithm called HFEL has been proposed in~\cite{Luo20}, which can approximate solutions to the device assignment problem. However, HFEL needs a long running time due to its search mechanism. Furthermore, if device scheduling is adopted in HFL, HFEL will be executed at each global iteration, thus leading to higher latency. Thus, there is a need to develop a device assignment algorithm that can efficiently assign IoT devices to edge servers while requiring less computation time than the HFEL algorithm.

In this paper, we propose an HFL framework to solve the two challenges above. Specifically, we design an improved K-Center (IKC) algorithm to perform device scheduling, which enables HFL to converge swiftly on \mbox{non-IID} datasets. In addition, we employ a deep reinforcement learning (DRL) model to tackle the device assignment problem. The well-trained DRL model performs comparably to the existing device assignment approach while achieving a faster-assigning process. The main contribution of this paper is summarized as follows:

\begin{enumerate}

\item We formulate a joint communication and computation optimization problem to minimize the weighted sum of time delay and energy consumption for training the entire HFL algorithm in an IoT system. We decompose the optimization problem into three subproblems: device scheduling problem, device assignment problem, and resource allocation problem.

\item We propose a vanilla K-Center (VKC) algorithm to deal with the device scheduling problem in HFL. We analyze the motivation behind VKC and reveal the flaws of VKC. Then, we propose an improved K-Center algorithm (IKC) to overcome these flaws.

\item We propose a dueling double deep Q-Network (D$^3$QN)-based device assignment algorithm for HFL. We carefully design the state space, action space, and reward function of D$^3$QN. We adopt bidirectional long short-term memory networks (BiLSTMs) as the D$^3$QN agent. We provide the workflow of training D$^3$QN in an HFL framework. After device scheduling and assignment, we adopt convex optimization tools to perform resource allocation for each edge server.

\item We conduct extensive numerical experiments to evaluate the proposed methods. IKC enables HFL to select only a fraction of IoT devices for model training while ensuring learning efficiency. D$^3$QN balances the workloads across the edge servers without causing high assigning latency. The proposed HFL framework significantly reduces the system cost compared with the baseline methods.

\end{enumerate}

\section{Related Work\\[0em]}\label{section2}
Federated Learning (FL) has attracted great attention from researchers because of its potential to address privacy concerns and improve the scalability of distributed machine learning (ML)~\cite{yang2022lead,Liu22,guo2021privacy}. To further enhance the efficiency of Federated Learning, a variety of resource allocation schemes are proposed to minimize FL's training time or energy consumption, or both~\cite{Yang20,Yang21,Dinh20}. In HFL, each edge server can apply these resource allocation methods individually. Therefore, resource allocation in HFL is not the core of our work. In the realm of device scheduling for FL, several studies have proposed various strategies. \cite{Shi21} introduced a strategy that combines device scheduling and resource allocation to maximize model accuracy within latency-constrained wireless FL settings. \cite{Amiri20} proposes a device scheduling algorithm that takes into account both channel conditions and the importance of local model updates. \cite{Ren20} designs a scheduling policy aimed at striking a balance between channel quality and update significance. \cite{McMahan17} introduces FedAvg for FL, which randomly schedules a subset of IoT devices during each iteration. \cite{zhang2023deep} utilizes deep reinforcement learning to minimize overall system costs through device scheduling. At each global iteration, however, \cite{Shi21,Amiri20,Ren20,zhang2023deep} require to collect model parameters or loss gradients from all IoT devices to the cloud for making scheduling decisions. In HFL, the cloud server only receives the parameters of the edge models at each global iteration and has no access to the parameters or the gradients of the local models. Therefore, these works are no longer effective in HFL. \cite{Xu21} presents a framework that combines device scheduling and resource allocation to enhance the long-term performance of FL. \cite{Zeng20Sky} proposes a design that integrates resource allocation and scheduling to improve the convergence rate of FL, all while adhering to energy and latency constraints. \cite{Zeng20} devises strategies for energy-efficient bandwidth allocation and scheduling, aiming to minimize total energy consumption without compromising learning performance. However, the bandwidth constraints considered in \cite{Xu21,Zeng20Sky,Zeng20} are intended for scenarios involving a single central server. In contrast, HFL has multiple edge servers with varying bandwidth resources. Given this context, these aforementioned approaches cannot be feasibly applied to HFL.

All the aforementioned works aim to enhance FL. In terms of HFL, \cite{Liu20} proposes a basic HFL framework and theoretically analyzes the convergence rate of HFL. \cite{Zhou23} proposes a detection mechanism for HFL to filter adverse devices and discusses the convergence of the proposed method. \cite{Zou23} adopts HFL to predict the day-ahead energy requirement of urban prosumers for the UAV system. \cite{Mohammadsadeq23} applies semi-synchronous communications to reduce the communication cost of the HFL system. However, device scheduling and device allocation are not discussed in \cite{Liu20,Zhou23,Zou23,Mohammadsadeq23}. \cite{Mhaisen21} proposes a device assignment algorithm for HFL to minimize the communication rounds while ensuring the accuracy of the model. \cite{Mhaisen21} ignores the time delay and energy consumption of HFL. \cite{zhang2023user} designs an HFL framework containing device assignment and resource allocation methods to decrease the energy cost and training latency. \cite{Luo20} proposes HFEL that solves both jointly solve the resource allocation and device assignment problems. However, the device assignment algorithms proposed in \cite{Luo20,zhang2023user} suffer from long assigning latency due to the large search space.

In contrast to prior research endeavors, the presented device scheduling methodology is specifically designed to cater to the specific requirements of HFL. This approach yields substantial reductions in both energy consumption and time delay. Furthermore, the proposed device assignment technique outperforms established benchmarks by expediting the assignment process while maintaining comparable performance.
\vspace{-1em}

\section{System Model}\label{section3} 
Consider an HFL system comprising $N$ IoT devices, denoted by the set $\mathcal{N} = \left\{1,2,...,N\right\}$, alongside $M$ edge servers, designated by the set $\mathcal{M} = \left\{1,2,...,M\right\}$, in addition to a central cloud server. Each IoT device, indexed by $n$, possesses a local dataset represented as $\mathcal{D}_n$, containing a quantity of data samples denoted as $D_n$. The machine learning models that undergo training procedures on the IoT devices are referred to as "local models," while those subject to aggregation processes at the edge servers and the cloud server are respectively termed "edge models" and "global models."

\subsection{HFL training}
Let $\bm{w}^i$ be the global model's parameters during the $i$-th global iteration. The training process of HFL at the $i$-th global iteration is explained as follows.

1) \textbf{Device Scheduling and Assignment}: The cloud server schedules a subset $\mathcal{H}_i \in \mathcal{N}$ with $H_i$ IoT devices to participate the training process. This paper assumes that the number of scheduled devices is fixed as $H$ (i.e., $H_1 = ... = H_I = H$). Then, each selected device will be assigned to an edge server. Define $\Psi_i = \left\{\mathcal{N}_{1,i},..., \mathcal{N}_{M,i}\right\}$ as a device assignment pattern at the $i$-th global iteration, where $\mathcal{N}_{m,i}\ (m\in\mathcal{M})$ denotes the group of IoT devices assigned to edge server $m$, and $\mathcal{H}_i = \bigcup_{m \in \mathcal{M}}\mathcal{N}_{m,i}$. Finally, the cloud server broadcasts the global model $\bm{w}^i$ to all the edge servers. Define $\bm{w}_m^{i,q}$ as the parameters of edge model $m$ at the $q$-th edge iteration, and $\bm{w}_m^{i,0} = \bm{w}^i$. 

2) \textbf{Local Training}: At the onset of the $q$-th edge iteration during the $i$-th global iteration, IoT device $n\in\mathcal{N}_{m,i}$ fetches the parameters $\bm{w}_m^{i,q}$ from edge server $m$. We define $\bm{\delta}_n^{i,q,l}$ to represent the parameters of local model $n$ during the $l$-th local iteration within the $q$-th edge iteration, and $\bm{\delta}_n^{i,q,0}$ is initialized as $\bm{w}_m^{i,q}$. Local training employs a gradient descent technique to minimize the loss function in the following manner:
% At the beginning of the $q$-th edge iteration of the $i$-th global iteration, IoT device $n\in\mathcal{N}_{m,i}$ downloads $\bm{w}_m^{i,q}$ from edge server $m$. Define $\bm{\delta}_n^{i,q,l}$ as the parameters of local model $n$ at the $l$-th local iteration  of the $q$-th edge iteration, and $\bm{\delta}_n^{i,q,0} = \bm{w}_m^{i,q}$. Local training adopts a gradient descent algorithm to minimize the loss function as follows:
\begin{equation}
    \bm{\delta}_n^{i,q,l+1} = \bm{\delta}_n^{i,q,l} - \beta \cdot \nabla \Gamma_n(\bm{\delta}_n^{i,q,l}),
\label{localtrain}
\end{equation}
where $\nabla$ means computing the gradient with respect to the parameter, $\Gamma_n(\bm{\delta}_n^{i,q,l})$ represents  the loss function of local model $n$, $l$ is the index of the local iteration, and $\beta$ denotes the learning rate. Local training finishes upon reaching the predefined maximum local iterations, denoted as $L$.

3) \textbf{Edge Aggregation}: After $L$ local iterations, the selected IoT devices upload the local models to the corresponding edge servers. Each edge conducts aggregation by averaging the local models:
\begin{equation}
    \bm{w}_m^{i,q+1} = \frac{\sum_{n \in \mathcal{N}_{m,i}}D_n\bm{\delta}_n^{i,q,L}}{D_{\mathcal{N}_{m,i}}},
\label{edgeaggr}
\end{equation}
where $D_{\mathcal{N}_{m,i}}$ is represented as $\sum_{n\in\mathcal{N}_{m,i}}D_n$. The process of local training and edge aggregation continues iteratively until the maximum number of edge iterations $Q$ is reached.

4) \textbf{Cloud Aggregation}: After $Q$ edge iterations, the edge servers transmit their respective edge models to the cloud server, which subsequently combines them in the following manner:
\begin{equation}
    \bm{w}^{i+1} = \frac{\sum_{m=1}^MD_{\mathcal{N}_{m,i}} \bm{w}_m^{i,Q}}{D},
\label{cloudaggr}
\end{equation}
where $D$ is denoted as $\sum_{m\in\mathcal{M}}D_{\mathcal{N}_{m,i}}$. It can be noted that each edge iteration contains $L$ local iterations, and each global iteration contains $Q$ edge iterations. As a result, each global iteration contains $Q\times L$ local iterations.

Algorithm~\ref{HFL_training} provides the workflow of the aforementioned process. Existing studies have demonstrated that HFL enhances energy efficiency and alleviates network congestion compared with traditional FL~\cite{Liu20,Zhou23,Zou23,Mohammadsadeq23,Luo20}.

\begin{figure}[t]
\begin{algorithm}[H]
\textsl{}\setstretch{1}
\caption{HFL training at the $i$-th global iteration.}  
\label{HFL_training}
\begin{algorithmic}[H]
\Require{Set of edge servers $\mathcal{M} = \{1,...,M\}$, set of scheduled IoT devices $\mathcal{H}_i$, global model $\bm{w}^{i}$, device assignment pattern $\Psi_i = \{\mathcal{N}_{1,i},...,\mathcal{N}_{M,i}\}$, maximum local iteration $L$, maximum edge iteration $Q$}
\Ensure{Global model $\bm{w}^{i+1}$}
\State The cloud server broadcasts $\bm{w}^i$ to the edge servers
\For{$q = 1$ to $Q$}
    \For{each IoT device $n \in \mathcal{H}_i$ in parallel}
        \State Download $\bm{w}^{i,q}_m$ from the corresponding edge server
	    \For{$l = 1$ to $L$}
			\State Conduct local training according to~\eqref{localtrain}
		\EndFor
		\State Upload $\bm{\delta}_n^{i,q,L+1}$ to the corresponding edge server
	\EndFor
	\For{each edge server $m\in\mathcal{M}$ in parallel}
		\State Conduct edge aggregation according to~\eqref{edgeaggr} to obtain $\bm{w}_m^{i,q+1}$
	\EndFor
\EndFor
\State The edge servers transmit the edge models to the cloud server
\State The cloud server conducts cloud aggregation according to~\eqref{cloudaggr} and derive the updated parameters $\bm{w}^{i+1}$
\State \Return $\bm{w}^{i+1}$
\end{algorithmic}
\end{algorithm}
\vspace{-2em}
\end{figure}

\begin{figure*}[htb]
  \centering
  \includegraphics[width=15cm]{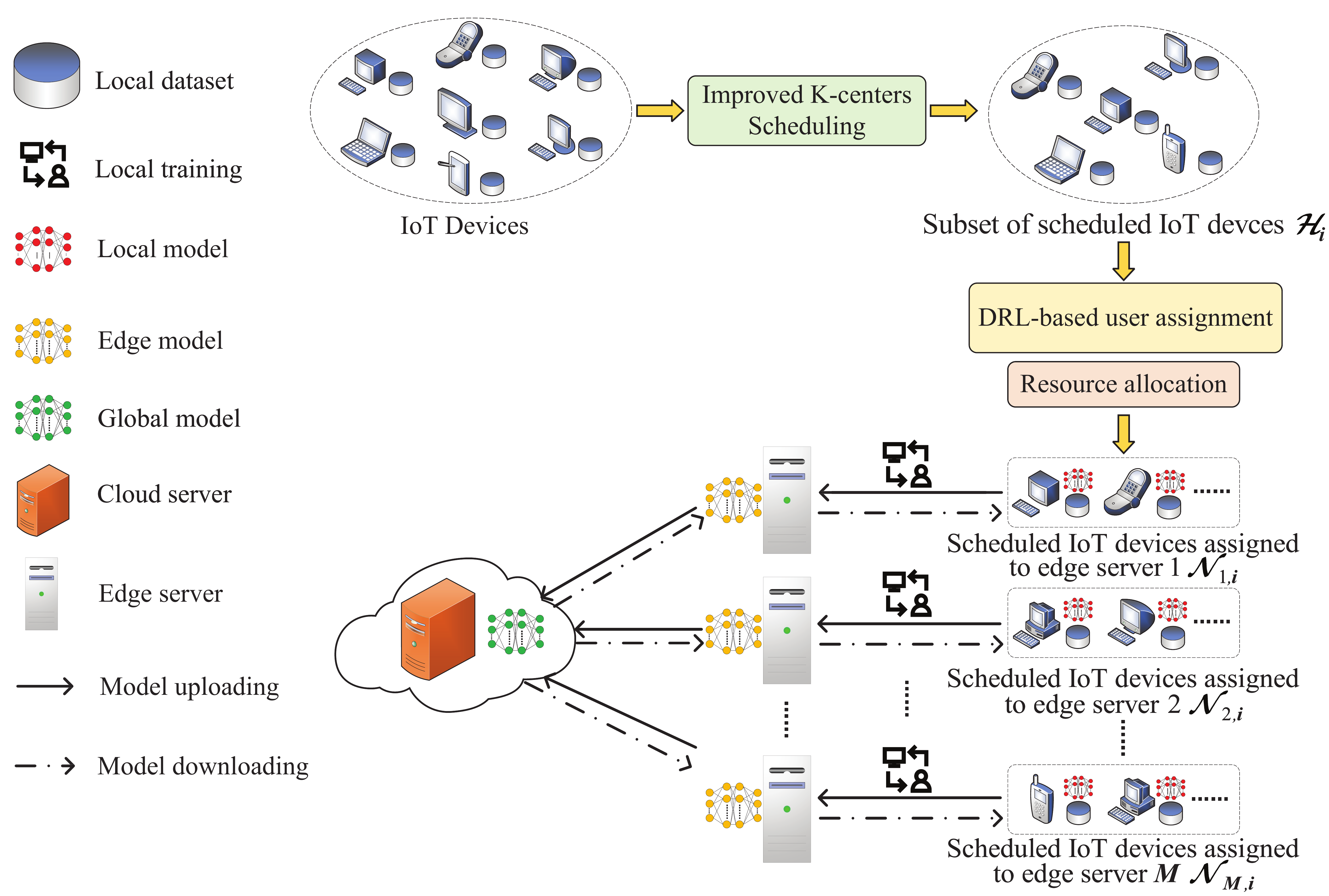}
\caption{Overview of the proposed HFL framework. At the beginning of the $i$-th global iteration, the improved K-centers scheduling (IKC) method is carried out to form the set $\mathcal{H}_i$. Next, the scheduled devices are assigned to the corresponding edge servers via the DRL-based device assignment method. Then, each edge server performs resource allocation to determine the bandwidth and CPU frequency of the scheduled devices. Finally, the training process begins according to Algorithm~\ref{HFL_training}. }
\label{FL_Framework}
\vspace{-1em}
\end{figure*}

\subsection{Energy consumption and time delay in HFL}\label{EnergyLatencyModel}
Analyzing data computation and transmission processes allows for the determination of HFL's energy consumption and time delay. $u_n$ denotes the number of CPU cycles requisite for the computation of a single data sample by IoT device $n$. IoT Device $n$ operates at CPU frequency $f_n$. After each edge iteration, the energy consumption $E_{n,i}^{\text{cmp}}$ and time delay $T_{n,i}^{\text{cmp}}$ of IoT device $n$ for processing $D_n$ data samples are:
\begin{equation}
    T_{n,i}^{\text{cmp}}= \frac{Lu_nD_n}{f_n},\ n\in\mathcal{H}_i,
\label{time_cmp}
\end{equation}
\begin{equation}
    E_{n,i}^{\text{cmp}} = \frac{\alpha}{2}L f_n^2 u_n D_n,\ n\in\mathcal{H}_i,
\label{energy_cmp}
\end{equation}
where $\frac{\alpha}{2}$ means the effective capacitance coefficient of the device's computing chipset.

When the IoT devices finish $L$ local iterations, the IoT devices send the parameters to their respective edge server. In this work, we adopt a frequency-division multiple access (FDMA) protocol for data transmission. During model uploading, IoT device $n$'s transmission rate will be calculated as follows:
\begin{equation}
    \eta_n = b_n\text{log}_2(1+\frac{\bar{g}^m_n p_n}{N_0b_n}),\ n\in\mathcal{H}_i,
\end{equation}
where $b_n$ represents the bandwidth allocated to device $n$, $\bar{g}^m_n$ means the average channel gain between IoT device $n$ and edge server $m$ during the entire training phase, $N_0$ denotes the background noise, and $p_n$ represents the transmit power. At each global iteration, the energy consumption $E_{n,i}^{t}$ and the time delay $T_{n,i}^t$ associated with the transmission of model parameters from IoT device $n$ to the edge can be described as follows:
\begin{equation}
    T^{\text{com}}_{n,i} = \frac{z}{\eta_n},\ n\in\mathcal{H}_i,
\end{equation}
\begin{equation}
    E^{\text{com}}_{n,i} = p_{n}T^{\text{com}}_{n,i},\ n\in\mathcal{H}_i,
\end{equation}
where $z$ is the size of the model parameters. Note that the latency and energy cost of downloading the aggregated model are typically considered negligible in existing works as the server possesses a notably higher average transmit power compared with the IoT devices, enabling them to effectively utilize the complete downlink bandwidth when transmitting the model~\cite{Xu21,Yang20,Yang21,Dinh20,Ren20,Shi21,Amiri20}. Thus, the time delay $T^\text{edge}_{m,i}$ and energy consumption $E^\text{edge}_{m,i}$ of the devices assigned to edge server $m$ for finishing $Q$ edge iterations can be derived as follows:
\begin{equation}
    T^\text{edge}_{m,i} = Q\max_{n \in \mathcal{N}_{m,i}}\left\{T_{n,i}^{\text{cmp}} + T^{\text{com}}_{n,i} \right\},\ m\in\mathcal{M},
\label{Tm}
\end{equation}
\begin{equation}
    E^\text{edge}_{m,i} = Q\sum_{n \in \mathcal{N}_{m,i}}(E_{n,i}^{\text{cmp}} + E^{\text{com}}_{n,i}),\ m\in\mathcal{M}.
\label{Em}
\end{equation}

The energy consumption denoted as $E^{\text{cloud}}_{m,i}$ and the time delay represented as $T^{\text{cloud}}_{m,i}$ for uploading edge models from edge server $m$ to the cloud server after $Q$ edge iterations can be computed as follows:
\begin{equation}
    T^{\text{cloud}}_{m,i} = \frac{z}{B\text{log}_2(1+\frac{\bar{g}^\text{cloud}_m p^m}{N_0 B})},\ m\in\mathcal{M}
\end{equation}
\begin{equation}
    E^{\text{cloud}}_{m,i} = p^mT^{\text{cloud}}_{m,i}, \ m\in\mathcal{M}
\end{equation}
where $\bar{g}^\text{cloud}_m$ is the mean channel gain observed between edge server $m$ and the cloud throughout the entire training period, $p^m$ is the transmission power of edge server $m$ during data transmission, and $B$ is the bandwidth allocated to the edge servers. In this paper, we assume that the cloud allocates equal bandwidth to each edge server; $p^m$ and $B$ remain static during the training process. As a result, $T^{\text{cloud}}_{m,i}$ and $E^{\text{cloud}}_{m,i}$ are constant parameters. Finally, the total energy consumption $E$ and time delay $T$ of training the entire HFL algorithm are obtained as
\begin{equation}
    T_{m,i} = T^{\text{cloud}}_{m,i}+T^\text{edge}_{m,i}, T_i =\max_{m \in \mathcal{M}}\left\{T_{m,i}\right\},  T = \sum_{i=1}^I T_i,
\end{equation}
\begin{equation}
    E_{m,i} = E^{\text{cloud}}_{m,i} + E^\text{edge}_{m,i}, E_i = \sum_{m \in \mathcal{M}} E_{m,i}, E = \sum_{i=1}^I E_i,
    \label{E_Total}
\end{equation}
where $I$ is the index of the global iteration, $T_{m,i}$ and $E_{m,i}$ respectively stand for the time delay and energy consumption associated with edge server $m$ during the $i$-th global iteration. Additionally, we have $T_i$ and $E_i$ representing the time delay and energy consumption for training HFL during the $i$-th global iteration."

\subsection{Problem formulation}
This paper introduces an optimization problem aimed at minimizing the weighted sum of the time delay and energy consumption for training the entire HFL algorithm:
\begin{align}
&\underset{\bm{\mathcal{B}}, \bm{\mathcal{F}},\bm{\mathcal{H}}, \bm{\Psi}}{\text{min}}\quad E + \lambda T  \label{YY}\\
&\textrm{s.t.} \ \; \sum_{n \in \mathcal{N}_{m,i}}b_n\leq B_m,~\forall m \in \mathcal{M}, \tag{\ref{YY}{a}} \label{YYa}\\
&\quad\ \;\; \,0 \leq f_n \leq f^{\text{max}}_n,~\forall n \in \mathcal{N}, \tag{\ref{YY}{b}} \label{YYb}\\
&\quad\ \;\; \,A_i < A^\text{target},\ i = 1,..., I-1, \tag{\ref{YY}{c}} \label{YYc}\\
&\quad\ \;\; \,A_I \geq A^\text{target},\tag{\ref{YY}{d}} \label{YYd}\\
&\quad\ \;\;\bigcup_{m \in \mathcal{M}}\mathcal{N}_{m,i} = \mathcal{H}_i,~\mathcal{H}_i \subseteq \mathcal{N},~i = 1,...,I, \tag{\ref{YY}{e}} \label{YYe} \\
&\quad\ \;\; \,\mathcal{N}_{\mu,i} \bigcap \mathcal{N}_{\nu,i} = \varnothing,~\forall \mu,\nu \in \mathcal{M},~\mu\neq\nu,~i = 1,...,I, \tag{\ref{YY}{f}} \label{YYf}
\end{align}
where $\lambda$ serves as a parameter that reflects the balance between $T$ and $E$, $\bm{\mathcal{B}} = \big[\bm{b}_i \big| i = 1,...,I  \big]$, $\bm{b}_i =  \big[b_n \big| n\in\mathcal{H}_i \big]$, $\bm{\mathcal{F}} =  \big[\bm{f}_i \big| i = 1,...,I  \big]$, $\bm{f}_i =  \big[f_n \big| n\in\mathcal{H}_i \big]$, $\bm{\mathcal{H}} = [\mathcal{H}_i \big| i = 1,...,I ]$, $\bm{\Psi} = [\Psi_i \big| i = 1,...,I ]$, $B_m$ is the total bandwidth of edge server $m$, $A^\text{target}$ is the target accuracy to indicate the convergence, $A_i$ is the accuracy derived by evaluating the global model on the testing set under the $i$-th global iteration,  $f^{\text{max}}_n$ represents the maximum CPU frequency of device $n$. \eqref{YYa} and \eqref{YYb} denote the bandwidth constraint and CPU frequency constraint, respectively. \eqref{YYc} and \eqref{YYd} indicate that the training process of HFL will be terminated when the global model achieves the target accuracy. \eqref{YYe} means that a subset of IoT devices $\mathcal{H}_i\in\mathcal{N}$ is selected to participate in the training process at each global iteration. \eqref{YYf} ensures that each selected IoT device communicates with only one edge server. Note that the value of $\lambda$ is determined based on the specific requirements of the practical scenarios. For example, if the project aims to achieve fast training speed, $\lambda$ will be set to a large value.

It is challenging to derive the globally optimal solution of optimization problem~\eqref{YY} as problem~\eqref{YY} couples device scheduling, device assignment, and spectrum resource allocation together, which involves multi-variable optimization and combinatorial optimization. This manuscript introduces an HFL framework illustrated in Fig.~\ref{FL_Framework} for deriving a locally optimal solution to problem~\eqref{YY}. First of all, improved K-Center (IKC) schedules the IoT devices at each global iteration. Then, a DRL-based device assignment algorithm derives the assignment pattern $\Psi_i$. Finally, a resource allocation approach is deployed for optimizing the allocated bandwidth and CPU frequency of IoT devices. IKC enables HFL to achieve target accuracy swiftly (i.e., minimizing $I$), and DRL-based device assignment together with the resource allocation algorithm aims to minimize the one-round system cost $E_i + \lambda T_i$. As a result, the proposed HFL framework effectively minimizes the total system cost $E+\lambda T$.

% \vspace{-0.5em}
\section{Improved K-Center for Device Scheduling}\label{device_sche}
In this section, we explain the adverse impact of \mbox{non-IID} datasets on training HFL. Then, we outline the motivation of designing a vanilla K-Centers (VKC) scheduling algorithm in HFL as well as its drawbacks. Finally, we propose an improved K-Centers (IKC) algorithm that addresses the shortcomings of the VKC algorithm.

\subsection{Vanilla K-Center}
In HFL, \mbox{non-IID} datasets usually refer to datasets that have highly skewed class distribution. For example, most of the data belong to a majority class, while the remaining data belong to other classes. The local model trained on the \mbox{non-IID} dataset achieves good performance in identifying its majority class while often failing to correctly predict the other classes~\cite{Alberto18}. Such performance bias of the local model can be transferred to the global model through edge aggregation and cloud aggregation. Take an HFL task trained on \mbox{non-IID} CIFAR-10\footnote{CIFAR-10 is a ten-class dataset~\cite{Krizhevsky09}. The ten different classes include airplane, autoIoT, bird, cat, deer, dog, frog, horse, ship, and truck.} as an example. If few of the scheduled local datasets contain samples from the ``bird'' class, the global model is more likely to make incorrect predictions when dealing with samples labeled as ``bird''.

Although scheduling fewer devices mitigates communication overheads and the straggler’s effect, HFL may require more global iterations to reach the convergence, especially when the local datasets are \mbox{non-IID}~\cite{Shi21}. To design an effective device scheduling algorithm, this work analyzes the relationship between the global model and the local datasets. According to~\eqref{localtrain},~\eqref{edgeaggr}, and~\eqref{cloudaggr}, the global model $\bm{w}^{i+1}$ can be rewritten as
\begin{align}
&\bm{w}^{i+1}  \notag\\
&=\frac{1}{D}\Big(\sum_{m=1}^M D_{\mathcal{N}_{m,i}}\frac{1}{D_{\mathcal{N}_{m,i}}}\sum_{n\in\mathcal{N}_{m,i}} D_n \bm{\delta}_n^{i,Q,L+1}   \Big) \notag \\
&=\frac{1}{D}\Big[\sum_{m=1}^M D_{\mathcal{N}_{m,i}}\frac{1}{D_{\mathcal{N}_{m,i}}}\sum_{n\in\mathcal{N}_{m,i}} \big( D_n \bm{\delta}_n^{i,Q,L} - \beta\nabla\Gamma_n(\bm{\delta}_n^{i,Q,L})\big)\Big] \notag \\
&= \frac{1}{D}\sum_{m=1}^M\sum_{n\in\mathcal{N}_{m,i}}\Big( D_n \bm{\delta}_n^{i,Q,L} - \beta\nabla\Gamma_n(\bm{\delta}_n^{i,Q,L}) \Big) \notag \\
&= \frac{1}{D}\sum_{n\in\mathcal{H}_{i}}\Big( D_n \bm{\delta}_n^{i,Q,L} - \beta\nabla\Gamma_n(\bm{\delta}_n^{i,Q,L}) \Big).
\label{WhyKCenter}
\end{align}
Define $\mathcal{D}_{\mathcal{H}_i}$ as the dataset containing all the samples that join the training process at the $i$-th global iteration. It can be noted from~\eqref{WhyKCenter} that the global model in HFL is essentially trained on $\mathcal{D}_{\mathcal{H}_i}$ in a distributed manner. Therefore, an IID $\mathcal{D}_{\mathcal{H}_i}$ is beneficial for the convergence of HFL even though each local dataset is still \mbox{non-IID}.

\begin{figure}[t]
\begin{algorithm}[H]
\textsl{}\setstretch{1}
\caption{K-means based device clustering}
\begin{algorithmic}[1] 
\Require{Set of IoT device $\mathcal{N}$, auxiliary model $\bm{w}^\text{aux}$}
\Ensure{$K$ clusters $\left\{\mathcal{C}_k|k = 1,...,K \right\}$}
\State Create $K$ empty sets $\left\{\mathcal{C}_k|k = 1,...,K \right\}$
\State The cloud server broadcasts auxiliary model $\bm{w}^\text{aux}$ to all the edge servers
\State Each IoT device is assigned to an edge server
\For{each IoT device $n \in \mathcal{N}$ \textbf{in parallel}}
    \State Download $\bm{w}^\text{aux}$ from the corresponding edge server
    \State Initialize the local auxiliary model $\bm{w}^\text{aux}_n(0) = \bm{w}^\text{aux}$
    \State Conduct local training based on~\eqref{localtrain} for $L$ iterations
    \State Transmit the model weights $\bm{w}^\text{aux}_n(L)$ to the cloud  server through the edge server
\EndFor
\State The cloud server trains a K-means model with $K$ clusters using $\{\bm{w}^\text{aux}_1(L),...,\bm{w}^\text{aux}_N(L)\}$

\For{each IoT device $n \in \mathcal{N}$}
    \State The K-means model provides the cluster label $k_n \in \{1,...,K\}$ for IoT device $n$
    \State $\mathcal{C}_{k_n} \leftarrow n$
\EndFor
\State \Return $\left\{\mathcal{C}_k|k = 1,...,K \right\}$
\end{algorithmic}  
\label{KMeans}
\end{algorithm}
\vspace{-3em}
\end{figure}

\begin{figure}[t]
\begin{algorithm}[H]
\textsl{}\setstretch{1}
\caption{Vanilla K-Center Scheduling (VKC)}
\begin{algorithmic}[1] 
\Require{Set of IoT devices $\mathcal{N}$, number of IoT devices selected from each cluster $h$}
\State The cloud server initializes the global model $\bm{w}^0$
\State Perform Algorithm~\ref{KMeans} while using $\bm{w}^0$ as the auxiliary model to obtain $K$ clusters $\left\{\mathcal{C}_k|k = 1,...,K \right\}$
\For{i = $1$ to $I$}
    \State Create an empty set $\mathcal{H}_i$
    \For{$k = 1$ to $K$}
        \If{$C_k \geq h$}
            \State Randomly schedule $h$ devices from $\mathcal{C}_k$ and add them to $\mathcal{H}_i$
        \Else
            \State Add all the devices from $\mathcal{C}_k$ to $\mathcal{H}_i$
        \EndIf
    \EndFor
    
    \If{$H_i < K \cdot h$}
        \State Obtain the set containing unscheduled devices $\overline{\mathcal{H}}_i = \mathcal{N} - (\mathcal{N}\bigcap\mathcal{H}_i)$
        \State Randomly schedule $(K \cdot h - H_i)$ devices from $\overline{\mathcal{H}}_i$ and add them $\mathcal{H}_i$
    \EndIf
    
    \State The IoT devices in $\mathcal{H}_i$ join HFL training based on Algorithm~\ref{HFL_training}
\EndFor
\end{algorithmic}  
\label{VKC}
\end{algorithm}
\vspace{-1em}
\end{figure}

\begin{figure}[t]
\begin{algorithm}[H]
\textsl{}\setstretch{1}
\caption{Improved K-centers (IKC)}  
\label{IKC}
\begin{algorithmic}[1] 
\Require{Set of IoT devices $\mathcal{N}$, number of IoT devices selected from each cluster $h$, mini model $\bm{\xi}$}
\State The cloud server initializes the global model $\bm{w}^0$ and mini model $\bm{\xi}$ 
\State Perform Algorithm~\ref{KMeans} while using $\bm{\xi}$ as the auxiliary model to obtain $K$ clusters $\left\{\mathcal{C}_k|k = 1,...,K \right\}$
\State Create $K$ empty sets $\mathcal{G}_1,...,\mathcal{G}_K$
\For{$i = 1$ to $I$}
    \State Create $K$ empty sets $\mathcal{H}^1_i,...,\mathcal{H}^K_i$
    \For{$k = 1$ to $K$}
        \If{$C_k + G_k \geq h$}
            \If{$C_k \geq h$}
                \State Randomly transfer $h$ devices from $\mathcal{C}_k$ to both $\mathcal{H}^k_i$ and $\mathcal{G}_k$
            \Else
                \State Transfer all the devices from $\mathcal{C}_k$ to $\mathcal{H}^k_i$
                \State Randomly transfer $(h-C_k)$ devices from $\mathcal{G}_k$ to $\mathcal{H}^k_i$
                \State Transfer the remaining devices from $\mathcal{G}_k$ to $\mathcal{C}_k$
                \State $\mathcal{G}_k$ copies the devices in $\mathcal{H}^k_i$
            \EndIf
        \Else
            \State Add all the devices from $\mathcal{C}_k$ to $\mathcal{H}^k_i$
        \EndIf
    \EndFor
    \State Combine $\mathcal{H}^1_i,...,\mathcal{H}^k_i$ together as $\mathcal{H}_i$
    
    \If{$H_i < K \cdot h$}
        \State Obtain the set containing unscheduled devices $\overline{\mathcal{H}}_i = \mathcal{N} - (\mathcal{N}\bigcap\mathcal{H}_i)$
        \State Randomly schedule $(K \cdot h - H_i)$ devices from $\overline{\mathcal{H}}_i$ and add them $\mathcal{H}_i$
    \EndIf
    \State The selected IoT devices $\mathcal{H}_i$ join HFL training based on Algorithm~\ref{HFL_training}
\EndFor
\end{algorithmic}
\end{algorithm}
\vspace{-0em}
\end{figure}

In this paper, we introduce a clustering-based scheduling method, so-called vanilla K-Center (VKC), for approximating $\mathcal{D}_{\mathcal{H}_i}$ to an IID dataset. First, the cloud server broadcasts an auxiliary model $\bm{w}^\text{aux}$ to all the IoT devices through the edge servers. Second, the auxiliary model at the $n$-th device (represented as $\bm{w}^\text{aux}_n$) is trained on $\mathcal{D}_n$. The trained $\bm{w}^\text{aux}_n$ is sent back to the cloud server through the corresponding edge server. Third, a K-means model is trained based on the parameters of the auxiliary models to divide the IoT devices into $K$ clusters. Generally, the number of $K$ is equal to the total number of classes of the dataset. In this case, the IoT devices whose datasets have the same majority class are divided into the same group due to the similar data distribution. Algorithm~\ref{KMeans} depicts the workflow of device clustering. Define $\mathcal{C}_k$ as the $k$-th cluster with $C_k$ IoT devices. Finally, at each global iteration, the same number of IoT devices are randomly scheduled from each cluster $\mathcal{C}_k$ to formulate the set $\mathcal{H}_i$, thus approximating $\mathcal{D}_{\mathcal{H}_i}$ to a balanced dataset. Define $h$ as the number of scheduled devices from each cluster. Algorithm~\ref{VKC} provides the details of implementing VKC in HFL. According to Lines 1-2, VKC uses the global model $\bm{w}^0$ as the auxiliary model $\bm{w}^\text{aux}$ to conduct device scheduling. If the number of the IoT devices in $\mathcal{C}_k$ exceeds $h$ (i.e., $C_k \geq h$), VKC randomly schedules $h$ different devices from $\mathcal{C}_k$ (Line 7). Otherwise, all the devices in $\mathcal{C}_k$ are added to $\mathcal{H}_i$ (Line 9). Then, VKC randomly schedules $(h\cdot K-H_i)$ devices from those that have not yet been scheduled at the current global iteration (Lines 13-14). Finally, the scheduled devices in $\mathcal{H}_i$ participate in the $i$-th global iteration.

% \vspace{-1em}

\subsection{Improved K-Center}\label{Section_IKC}
The deployment of VKC suffers from two main defects. First, the size of the HFL model (i.e., $z$) will be quite large if HFL adopts complex deep neural networks like Convolutional Neural Networks (CNNs). Besides, Algorithm~\ref{KMeans} involves all the IoT devices for training the auxiliary model $\bm{w}^\text{aux}$. As a result, VKC will cause excessive communication and computation overheads when performing device clustering. Second, it is possible that the same devices are repeatedly scheduled in successive iterations, while some devices are infrequently scheduled, resulting in inefficient utilization of the IoT devices' data.

In this paper, we propose improved K-centers (IKC) to address the limitations of VKC. Instead of using the heavyweight model $\bm{w}^0$, IKC adopts a mini model $\bm{\xi}$ that only contains one convolutional layer and one linear layer to carry out device clustering. To further reduce the size of the mini model $\bm{\xi}$, IKC conducts data preprocessing to shrink the input dimension of $\bm{\xi}$. Take an image classification task trained on CIFAR-10 as an example. The shape of the images in CIFAR-10 is $3\times32\times32$ (channels$\times$height$\times$width). To reduce the input dimensions, only one channel of the images is used for training the mini model $\bm{\xi}$, and the images are randomly cropped into $10\times 10$. For other data types, feature selection approaches like principal component analysis~\cite{Karl01} can be applied to reduce the data dimensions. As a result, IKC considerably reduces the computation and communication costs for performing Algorithm~\ref{KMeans}.

To address the issue caused by repetitive scheduling, IKC maintains a record of the IoT devices that have been scheduled in previous iterations and prioritizes unscheduled IoT devices. IKC significantly diversifies the data distribution of the training data $\mathcal{D}_{\mathcal{H}_i}$, thus enabling HFL to reach the target accuracy within fewer global iterations.

Algorithm~\ref{IKC} provides the details of IKC. To begin with, IKC employs mini model $\bm{\xi}$ to conduct device clustering and initializes $K$ empty sets $\mathcal{G}_1,...,\mathcal{G}_K$ for recording the scheduled devices (Lines 1-3). Define $G_k$ as the number of devices in $\mathcal{G}_k$. If $C_k\geq h$, IKC will randomly transfer $h$ devices from $\mathcal{C}_k$ to $\mathcal{H}_i^k$, and $\mathcal{G}_k$ records the selected devices (Line 9). If $C_k < h$ and $C_k + G_k \geq h$, all the devices in $\mathcal{C}_k$ are transferred to $\mathcal{H}_i^k$, which makes $\mathcal{C}_k$ an empty set (Line 11). Next, ($h-C_k$) devices are randomly transferred from $\mathcal{G}_k$ to $\mathcal{H}_i^k$ (Line 12). Then, $\mathcal{G}_k$ is emptied by transferring all the remaining devices to $\mathcal{C}_k$ (Line 13). Finally, $\mathcal{G}_k$ copies the device indices in $\mathcal{H}_i^k$ (Line 14). If $C_k + G_k < h$, IKC will schedule the IoT devices in the same way as Lines 13-14 of Algorithm~\ref{VKC} (Lines 22-23). 

\section{Fast Device Assignment via Deep Reinforcement Learning}\label{DRL_assignment}

\subsection{Preliminaries\\[0em]}\label{preliminaries}

Given the set of scheduled devices $\mathcal{H}_i$, the device assignment problem aims to minimize the one-round weighted sum as follows
\begin{align}
&\underset{\bm{b}_i, \bm{f}_i, \Psi_i}{\text{min}}\quad E_i + \lambda T_i  \label{XX}\\
&\textrm{s.t.} \ \; \sum_{n \in \mathcal{N}_{m,i}}b_n\leq B_m,~\forall m \in \mathcal{M}, \tag{\ref{XX}{a}} \label{XXa}\\
&\quad\ \;\; \,0 \leq f_n \leq f^{\text{max}}_n,~\forall n \in \mathcal{N}, \tag{\ref{XX}{b}} \label{XXb}\\
&\quad\ \;\; \,\mathcal{N}_{\mu,i} \bigcap \mathcal{N}_{\nu,i} = \varnothing,~\forall \mu,\nu \in \mathcal{M},~\mu\neq\nu,~i = 1,...,I, \tag{\ref{XX}{c}} \label{XXd}
\end{align}

It can be observed that problem~\eqref{XX} is essentially a combinatorial optimization problem coupled with a multi-variable optimization problem. A recent study proposes an iterative algorithm called HFEL to deal with problem~\eqref{XX}. Firstly, HFEL performs device transferring adjustment, which means transferring a device from its original edge server to another server. Secondly, HFEL carries out device exchanging adjustments, which will exchange two devices between their respective edge servers. HFEL will perform resource allocation algorithms for each edge server and obtain the objective value~\eqref{XX} before and after each adjustment. The adjustments are permitted if they can reduce the objective value~\eqref{XX}. HFEL iteratively conducts device transferring adjustments and device exchanging adjustments until no more adjustments are permitted.

It is possible for HFEL to converge to the global optimal solution of problem~\eqref{XX} after a large number of iterations. However, due to the large search space, HFEL incurs high assigning latency in reaching convergence. Moreover, HFEL assumes that all the devices are involved at each global iteration, in which case device assignment only needs to be carried out once. In this work, however, the devices that participate in training HFL vary at each global iteration due to device scheduling, thus necessitating device assignment at each global iteration. Therefore, HFEL will incur even higher assigning latency. 

\subsection{Dueling Double Deep Q-Network}\label{D3QN}
To overcome the shortcoming of HFEL, this paper adopts deep reinforcement learning (DRL) to approximate the performance of HFEL with a fast assigning speed.

DRL aims to solve the optimization problem that can be formulated as a Markov Decision Process (MDP). In a DRL environment, the characteristics of the environment at the $t$-th time slot are reflected by state $\bm{s}_t$. The agent, which is usually a neural network with parameters $\theta$, takes action $a_t$ and obtains an immediate reward $r_t$ after receiving $\bm{s}_t$. The goal of DRL is to maximize the discounted accumulated reward $R_t=\sum_{j = 0}^{\infty}\gamma^j r_{t+j}$. In addition, DRL defines the state-action value function $Q(\bm{s}, a;\theta)$ and state-value function $V(\bm{s};\theta)$ as follows
\begin{equation}
    Q(\bm{s}, a; \theta) = \mathbb{E}\big[R_t \big| \bm{s}_t = \bm{s}, a_t = a; \theta \big],
    \label{Q_function}
\end{equation}
\begin{equation}
    V(\bm{s};\theta) = \mathbb{E}_{a\sim\pi(\bm{s})}\left[Q^\pi(\bm{s}, a; \theta)\right].
    \label{V_function}
\end{equation}

To tackle device assignment through DRL, the agent is designed to assign one IoT device to an edge server at each time slot. In this case, an episode of DRL will contain $H$ time slots if $H$ IoT devices are scheduled at each global iteration (i.e., $t=1,...,H$). Define $n_t$ as the IoT device that will be assigned at the $t$-th time slot. Besides, the assigning decision is the index of the edge server, and thus the action space of DRL is discrete. This paper employs Dueling Double Deep Q-network (D$^3$QN) to handle the discrete optimization problem. D$^3$QN merges the merits of the target network and dueling technology~\cite{Hasselt16,Ziyu16,Iqbal21}. D$^3$QN consists of two neural networks with identical architectures, known as the online network and the target network. The neural networks contain two separate heads. Define $\phi$ as the parameters of the shared layers in D$^3$QN. One head with parameters $\rho$ estimates the state-value function $V(\bm{s}_t;\phi, \rho)$, and the other head with parameters $\zeta$ estimates the advantage of action $a_t$ in state $\bm{s}$ $A(\bm{s},a;\phi, \zeta)$ (so called advantage function). Note that the parameters of the agent $\theta$ consists of $\phi$, $\rho$, and $\zeta$. Then, D$^3$QN calculates the Q-value as
\begin{align}
Q(\bm{s},a;\theta) = V(\bm{s};\phi , \rho )+ & (A(\bm{s},a;\phi, \zeta) -\frac{1}{M}\sum_{\bar{a}\in\mathcal{M}}A(\bm{s},\bar{a};\phi, \zeta)).
\label{Q_function_dueling}
\end{align}
D$^3$QN applies a replay buffer $\bm{\Omega}$ to store the tuple ($\bm{s}_t, a_t, r_t, \bm{s}_{t+1}$). Define $\big|\bm{\Omega}\big|$ as the size of the replay buffer. A minibatch $\mathcal{O}$ containing $O$ tuples are sampled from the replay buffer $\bm{\Omega}$ when training the DRL model $\pi_\theta$ using gradient descent. The loss function $L(\theta)$ is defined as
\begin{align}
    L(\theta) = 
    \frac{1}{O}\sum_{t'\in\mathcal{O}}\left(Q^\text{target}_{t'} - Q(\bm{s}_{t'},a_{t'};\theta) \right)^2,
    \label{loss_DRL}
\end{align}
where $Q^\text{target}_{t'}$ is derived as
\begin{align}
  Q^\text{target}_{t'}= \qquad\qquad\qquad\qquad\qquad\qquad\qquad\qquad\qquad\qquad\quad &&\notag\\\left\{\begin{aligned}
   &r_{t'}, &\text{if $\bm{s}_{t'}$ is terminal},  \\
   &r_{t'} + \gamma \underset{a_{t'+1}\in\mathcal{M}}{\text{max}}Q(\bm{s}_{t'+1}, a_{t'+1};\theta^\text{target}),& \text{otherwise},
  \end{aligned}\right. 
\label{target_cal}
\end{align}
where $\theta^\text{target}$ denotes the parameters of the target network. The online network $\theta$ is updated at each time step, while the target network $\theta^\text{target}$ is updated at fixed intervals to provide a more stable target.

\vspace{-1.5em}

\subsection{DRL Environment in HFL}
We carefully design the DRL environment, including the network architecture of the DRL model, state space, action space, and reward function as follows.

\textbf{Action:} As mentioned in Section~\ref{D3QN}, the agent assigns IoT device $n_t$ to edge server $a_t\in\mathcal{M}$ at each time slot $t$. D$^3$QN generates $a_t$ based on the Q-value:
\begin{equation}
    a_t = \underset{a\in\mathcal{M}}{\text{argmax}}Q(\bm{s}_t, a;\theta)
    \label{a_choose}
\end{equation}

\begin{figure}[t]
  \centering
  \includegraphics[width=1\linewidth]{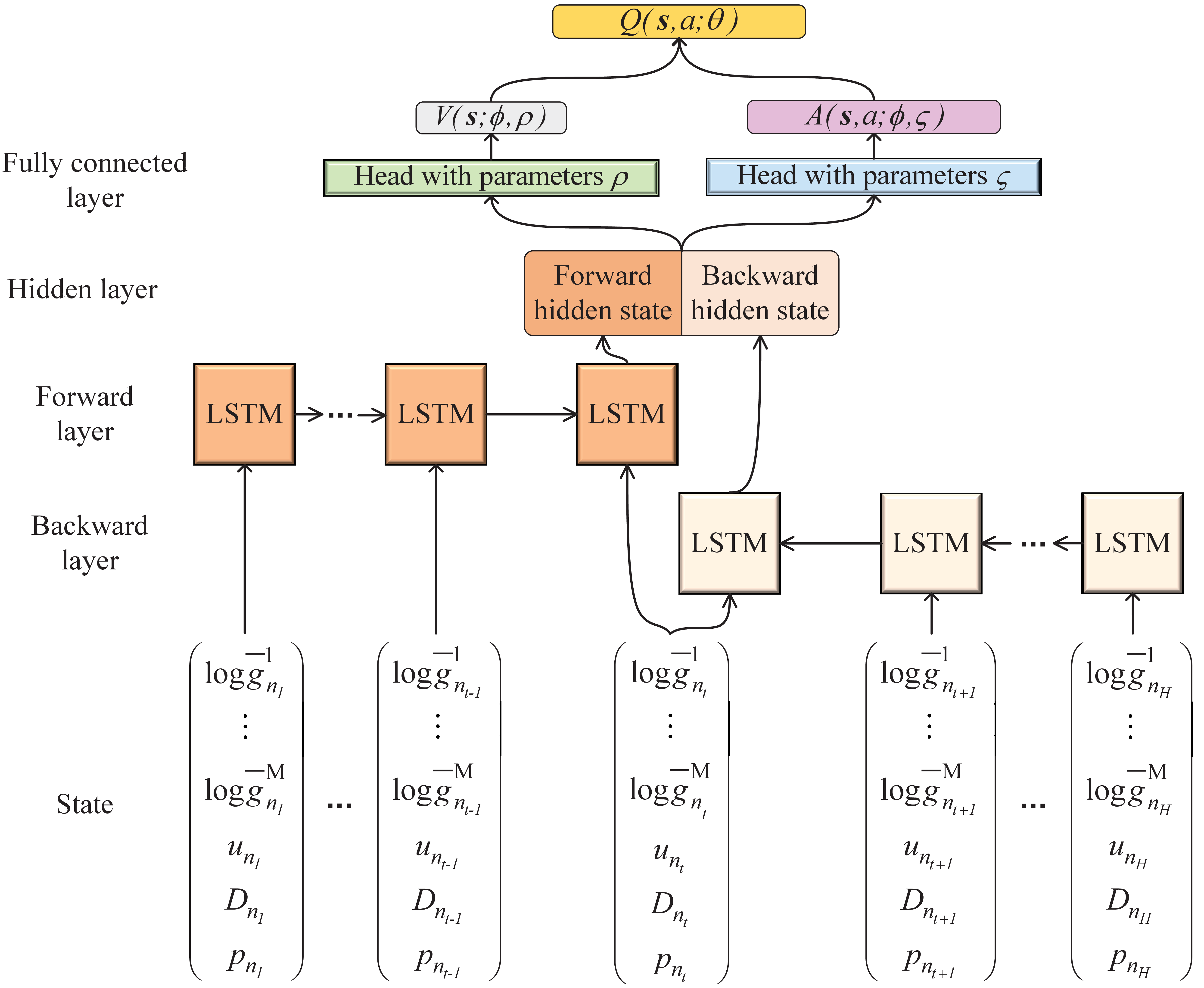}
\caption{Workflow of the BiLSTMs-based model at the $t$-th time slot. $\phi$ represents the parameters of the LSTM modules.\\[0em]}
\label{lstm_framework}
\end{figure}

\textbf{State and network architecture:} The DRL model needs to consider three aspects when assigning IoT devices $n_t$. First, the DRL model requires analyzing the channel states (e.g., $\bar{g}_{n_t}^m$) and device characteristics of device $n_t$ (e.g., $u_{n_t}$, $D_{n_t}$, etc.). Second, supposed device $n_t$ is assigned to edge server $m$ ($m\in\mathcal{M}$), it shares the bandwidth resources $B_m$ with the other devices that have already been assigned to the same server. Therefore, all the assigned devices should be considered when the DRL model is making an assignment decision $a_t$. Third, as the assignment decision of the current time slot will affect the subsequent assignment decisions, it is necessary for the DRL model to take the unassigned devices into consideration for maximizing long-term profits. Based on the above discussion, a simple idea of formulating state $\bm{s}_t$ is to use the channel states and device characteristics of all the scheduled devices $\big\{(\bar{g}_n^1, ..., \bar{g}_n^M, u_n, D_n, p_n)  \big| n\in\mathcal{H}_i\big\}$, and the DRL model is a multilayer perceptron (MLP) of which input and output dimensions are respectively $(M+3)\times H$ and $M$. Using the aforementioned method, however, DRL cannot be trained as the state information will remain the same when $t$ is from 1 to $H$. To address this issue, the network architecture of the agent in this paper adopts bidirectional long short-term memory networks (BiLSTMs). Moreover, we define $\bm{\chi}_{n_t}$ as the feature vector of device $n_t$ after min-max normalization:
\begin{equation}
    \bm{\chi}_{n_t} = (\widetilde{g}_{n_t}^1, ...,  \widetilde{g}_{n_t}^M, \widetilde{u}_{n_t}, \widetilde{D}_{n_t}, \widetilde{p}_{n_t}),
\end{equation}
where $\widetilde{g}_{n_t}^1$, ...,  $\widetilde{g}_{n_t}^M$, $\widetilde{u}_{n_t}$, $\widetilde{D}_{n_t}$, and $\widetilde{p}_{n_t}$ are normalized values. For example, $\widetilde{g}_{n_t}^1$ is derived by $\frac{\bar{g}_{n_t}^1 - \min\{\bar{g}_{n_t}^1\big| t = 1,...,H\}}{\max\{\bar{g}_{n_t}^1\big| t = 1,...,H\}-\min\{\bar{g}_{n_t}^1\big| t = 1,...,H\}}$. The channel states and device characteristics of the scheduled devices vary at different global iterations. Under this situation, min-max normalization enables the features to be presented in a fixed range. As a result, state $\bm{s}_t$ is constructed as
\begin{equation}
    \bm{s}_t = \big( (\underbrace{\bm{\chi}^\intercal_{n_1}, ..., \bm{\chi}^\intercal_{n_t}}_{\text{forward input}}), (\underbrace{ \bm{\chi}^\intercal_{n_t},..., \bm{\chi}^\intercal_{n_H}}_{\text{backward input}}) \big).
    \label{state}
\end{equation}
Note that in Equation~\eqref{state}, $\bm{\chi}^\intercal_{n_t}$ is used in both forward input and backward input. The device characteristics of the assigned and unassigned devices are included in the forward and backward input, respectively. Fig.~\ref{lstm_framework} depicts how the BiLSTM-based agent generates the Q-value after receiving $\bm{s}_t$.

\textbf{Reward:} The reward $r_t$ is defined as
\begin{equation}
  r_t=\begin{cases}
    1, & \text{if $n_t \in\widehat{\mathcal{N}}_{a_t}$},\\
    -1, & \text{otherwise},
  \end{cases}
\label{reward}
\end{equation}
where $\widehat{\mathcal{N}}_{a_t}$ is the set obtained by HFEL, which contains the devices assigned to the $a_t$-th edge server. According to ~\eqref{reward}, the DRL model is trained to make the same assignment decision as the HFEL method.

\begin{figure}[t]
\begin{algorithm}[H]
\textsl{}\setstretch{1}
\caption{Training process of the DRL model}
\begin{algorithmic}[1] 
\Require{Number of scheduled devices $H$, number of tuples in a minibatch $O$, update interval $J$, maximum episode $max\_ep$}
\Ensure{DRL model $\pi_\theta$}
\State Initialize the DRL model $\pi_\theta$ and replay buffer $\bm{\Omega}$
State Initialize the target network $\theta'$ with $\theta$
\State $ep = 0$, $step = 0$
\Repeat
    \State $\big\{(\bar{g}^1_{n_t}, ..., \bar{g}^M_{n_t}, u_{n_t}, D_{n_t}, p_{n_t} )\big| t = 1,...,H \big\}$ is randomly generated
    \State Obtain the assignment pattern $\widehat{\Psi} = \big\{\widehat{\mathcal{N}}_m \big| m\in \mathcal{M} \big\}$ via HFEL
    \For{$t = 1,...,H$}
        \State Formulate $\bm{s}_t$ based on~\eqref{state}
        \State Feed $\bm{s}_t$ to $\pi_\theta$ and obtain Q-values $\{Q(\bm{s}_t, a; \theta)\big|a\in\mathcal{A} \}$ based on~\eqref{Q_function_dueling}
        \State Select $a_t$ based on~\eqref{a_choose} and obtain $\bm{s}_{t+1}$
        \State Calculate $r_t$ based on~\eqref{reward}
        \State Push the tuple $(\bm{s}_t, a_t, r_t, \bm{s}_{t+1})$ into $\bm{\Omega}$
        \If{$\big|\bm{\Omega}\big|>O$}
            \State Randomly sample $O$ tuples from $\bm{\Omega}$
            \State Update $\theta$ by gradient descent with the loss function~\eqref{loss_DRL}
        \EndIf
        \State $step = step + 1$
        \If{$step \% J = 0$}
            \State Update the target network $\theta'$ with $\theta$
        \EndIf
    \EndFor
    \State $ep = ep + 1$
\Until $ep = max\_ep$
\State \Return $\pi_\theta$
\end{algorithmic}  
\label{TrainingDRL}
\end{algorithm}
\vspace{0em}
\end{figure}

\begin{figure}[t]
\begin{algorithm}[H]
\textsl{}\setstretch{1}
\caption{Workflow of the proposed HFL framework}  
\label{Overall_framework}
\begin{algorithmic}[1] 
\Require{Set of IoT devices $\mathcal{N} = \{1,...,N\}$, set of edge servers $\mathcal{M} = \{1,...,M\}$, preset accuracy $A^\text{target}$, number of IoT devices scheduled at each global iteration $H$}
\Ensure{Global model $\bm{w}$}
\State $i = 0$
\State The cloud server initializes global model $\bm{w}^{1}$
\Repeat
    \State $i = i+1$
    \State Schedule $H$ devices based on Algorithm~\ref{IKC}
    \State Obtain device assignment pattern $\Psi_i$ using the DRL model trained by Algorithm~\ref{TrainingDRL}
    \State Solve resource allocation problem~\eqref{ZZ} for each edge server using convex optimization 
    \Statex \quad\, tools like CVXPY
    % \Statex \;\;\;\;\,  based on Algorithm~\ref{ResourceAllocation}
    \State Perform model training to obtain $\bm{w}^{i+1}$ based on Algorithm~\ref{HFL_training}
    \State Evaluate $\bm{w}^{i+1}$ on the testing set to obtain the testing accuracy $A^i$
\Until{$A^i \geq A^\text{target}$}
\State $I = i$
\State \Return $\bm{w}^{I+1}$
\end{algorithmic}
\end{algorithm}
\vspace{-0em}
\end{figure}

\subsection{Resource allocation within a single edge server}

After obtaining $\Psi_i$, the resource allocation algorithm is conducted within each edge server to optimize the allocated bandwidth $\bm{b}_i$ and CPU frequency $\bm{f}_i$. Given edge server $m$ and its corresponding IoT devices $\mathcal{N}_{m,i}$, the resource allocation problem  is formulated as follows:
\begin{align}
&\underset{\bm{b}_{\mathcal{N}_{m,i}}, \bm{f}_{\mathcal{N}_{m,i}}}{\text{min}}\quad E_{m,i} + \lambda T_{m,i}  \label{ZZ}\\
&\textrm{s.t.} \ \; \sum_{n \in \mathcal{N}_{m,i}}b_n\leq B_m, \tag{\ref{ZZ}{a}} \label{ZZa}\\
&\quad\ \;\; \,0 \leq f_n \leq f^{\text{max}}_n,~\forall n \in \mathcal{N}_{m,i}, \tag{\ref{ZZ}{b}} \label{ZZb}
\end{align}
where $\bm{b}_{\mathcal{N}_{m,i}}=\{b_n \big| n\in\mathcal{N}_{m,i} \}, \bm{f}_{\mathcal{N}_{m,i}} = \{f_n \big| n\in\mathcal{N}_{m,i} \}$. 

According to~\eqref{time_cmp} to~\eqref{E_Total}, $E_{m,i} =  E^{\text{cloud}}_{m,i} + Q\sum_{n \in \mathcal{N}_{m,i}}( \frac{\alpha}{2}L f_n^2 u_n D_n + \frac{z p_{n}}{b_n\text{log}_2(1+\frac{\bar{g}^m_n p_n}{N_0b_n})})$ and $T_{m,i} =  T^{\text{cloud}}_{m,i} +  Q\max_{n \in \mathcal{N}_{m,i}}\Big\{ \frac{Lu_nD_n}{f_n}$ $+  \frac{z}{b_n\text{log}_2(1+\frac{\bar{g}^m_n p_n}{N_0b_n})} \Big\}$. Let $Y_n = \frac{\bar{g}^m_n p_n}{N_0}$, $\text{log}_2(1+\frac{Y_n}{b_n})$ is concave in $(b_n,Y_n)$ as $\text{log}_2(1+Y_n)$ is concave. Therefore, $b_n\text{log}_2(1+\frac{Y_n}{b_n})$ is concave. As the reciprocal of concave functions in $R^+$ is convex, $\frac{1}{b_n\text{log}_2(1+\frac{Y_n}{b_n})}$ is convex. In addition, $\frac{Lu_nD_n}{f_n}$ and $\frac{\alpha}{2}L f_n^2 u_n D_n$ are convex w.r.t. $f_n$. $E^{\text{cloud}}_{m,i}$ and $T^{\text{cloud}}_{m,i}$ are constant. Since the maximum of convex functions is also convex, the objective function~\eqref{ZZ} is convex. Besides, the constraints~\eqref{ZZa} and~\eqref{ZZb} are naturally convex.  Therefore, problem~\ref{ZZ} is a convex problem and can be solved by convex optimization tools like CVXPY~\cite{diamond2016cvxpy}.

\vspace{-0em}

\subsection{Workflow of Training the DRL Model}

Algorithm~\ref{TrainingDRL} provides the details of training the DRL model. Line 1 initializes the DRL model and creates an empty replay buffer. In Line 2, an integer variable $ep$ is introduced to trace the number of episodes. At the beginning of each episode (i.e., Lines 4-5), a set of channel states and device characteristics are randomly generated, and HFEL is used to obtain the device assignment baseline $\widehat{\Psi}$. In Lines 7-11, the DRL model produces $a_t$ and $r_t$ after receiving $\bm{s}_t$, and the tuple is stored into $\bm{\Omega}$. In Lines 13-14, $O$ tuples are randomly sampled from $\bm{\Omega}$, and the DRL model is trained by gradient descent using the loss function~\eqref{loss_DRL}. In Lines 16-19, the target network $\theta'$ is updated by the online network $\theta$ periodically. Finally, the training process will be terminated if the number of episodes is equal to the preset maximum episode.

Algorithm~\ref{Overall_framework} provides the interplay of the proposed device scheduling, device assignment, and resource allocation methods in an HFL framework.

\begin{table}[]
\centering
\caption{Parameters setup}
\textsl{}\setstretch{1}
\setlength{\tabcolsep}{1mm}{
\begin{tabular}{|c|c|}
\hline
\textbf{Parameter}                                                                                   & \textbf{Value}                                                                                                    \\ \hline
\begin{tabular}[c]{@{}c@{}}Number of CPU cycles to \\ process one sample $u_n$\end{tabular} & $[1,10]\times 10^4$ cycles/sample                                                                        \\ \hline
\begin{tabular}[c]{@{}c@{}}Total Bandwidth of edge \\ servers $B_m$\end{tabular}            & $[0.5,3] \text{MHz}$                                                                                     \\ \hline
\begin{tabular}[c]{@{}c@{}}Bandwidth allocated to edge \\ servers $B$\end{tabular}          & 10 MHz                                                                                                   \\ \hline
\begin{tabular}[c]{@{}c@{}}Average transmit power of\\ IoT devices $p_n$\end{tabular}      & $[0,23] \text{dBm}$                                                                                      \\ \hline
\begin{tabular}[c]{@{}c@{}}Average transmit power of\\ edge servers $p^m$\end{tabular}      & 23 dBm                                                                                                   \\ \hline
\begin{tabular}[c]{@{}c@{}}Maximum CPU frequency \\ $f^\text{max}$\end{tabular}             & 2 GHz                                                                                                    \\ \hline
Background noise $N_0$                                                                      & -174 dBm/MHz                                                                                             \\ \hline
\begin{tabular}[c]{@{}c@{}}Learning rate of HFL \\ $\beta$\end{tabular}                     & 0.01                                                                                                     \\ \hline
\begin{tabular}[c]{@{}c@{}}Maximum number of local\\ iterations $L$\end{tabular}            & 5                                                                                                        \\ \hline
\begin{tabular}[c]{@{}c@{}}Maximum number of edge\\ iterations $Q$\end{tabular}             & 5                                                                                                        \\ \hline
Number of clusters $K$                                                                      & 10                                                                                                       \\ \hline
Discounted factor $\gamma$                                                                  & 0.99                                                                                                     \\ \hline
Number of tuples $O$                                                                        & 128                                                                                                      \\ \hline
\begin{tabular}[c]{@{}c@{}}Size of mini model \\ $\bm{\xi}$\end{tabular}                    & 10 KB                                                                                                    \\ \hline
Size of the FL model $z$                                                                    & \begin{tabular}[c]{@{}c@{}}448 KB for FashionMNIST\\      882 KB for CIFAR-10\end{tabular}               \\ \hline
\begin{tabular}[c]{@{}c@{}}Size of local datasets \\ $D_n$\end{tabular}                     & \begin{tabular}[c]{@{}c@{}}{[}400,700{]} for FashionMNSIT\\      {[}300,600{]} for CIFAR-10\end{tabular} \\ \hline
\end{tabular}}
\label{para_setup}
\end{table}

\section{PERFORMANCE EVALUATION}\label{section7}
We consider $N = 100$ IoT devices and $M = 5$ edge devices randomly distributed within a square area with a one-kilometer side, and the center of the square is the cloud server. The path loss model is $128.1 + 37.6\log_{10}d$(km), where $d$ denotes the distance. The standard deviation of shadow fading is 8 dB. In terms of device assignment, the DRL model contains an LSTM module with 256 hidden units and two linear layers. The learning rate for training DRL is $10^{-3}$. 

Two public datasets, FashionMNIST~\cite{xiao2017} and CIFAR-10~\cite{Krizhevsky09}, are used to train the HFL model. FashionMNIST is a ten-class dataset containing 60000 training samples and 10000 testing samples. Each sample is a $1\times 28 \times 28$ gray-scale image. CIFAR-10 is a ten-class dataset consisting of 60000 $3\times 32 \times 32$ color images. The HFL model has two $5\times5$ convolutional layers and two linear layers. The output channels of the convolutional layers are 15 and 28, respectively. Each convolutional layer is followed by $2\times 2$ max pooling. In terms of mini model $\bm{\xi}$ in IKC, the input image is cropped to $1\times 10 \times 10$. The mini model $\bm{\xi}$ is constructed by a $2\times2$ convolutional layer and a linear layer, where the convolutional layer is followed by $2\times 2$ max pooling. The HFL models and mini model $\bm{\xi}$ are initialized by~\cite{He_2015_ICCV}. Unless otherwise specified, the parameters setup is listed in Table~\ref{para_setup}. 
\vspace{0em}

\subsection{Evaluation of device scheduling\\[-0em]}\label{device_result}

\begin{figure*}[t]
\centering
\hspace{-1em}
\subfigure[$H = 10$]{
\begin{minipage}[t]{0.25\linewidth}
\centering
\includegraphics[width=1\linewidth]{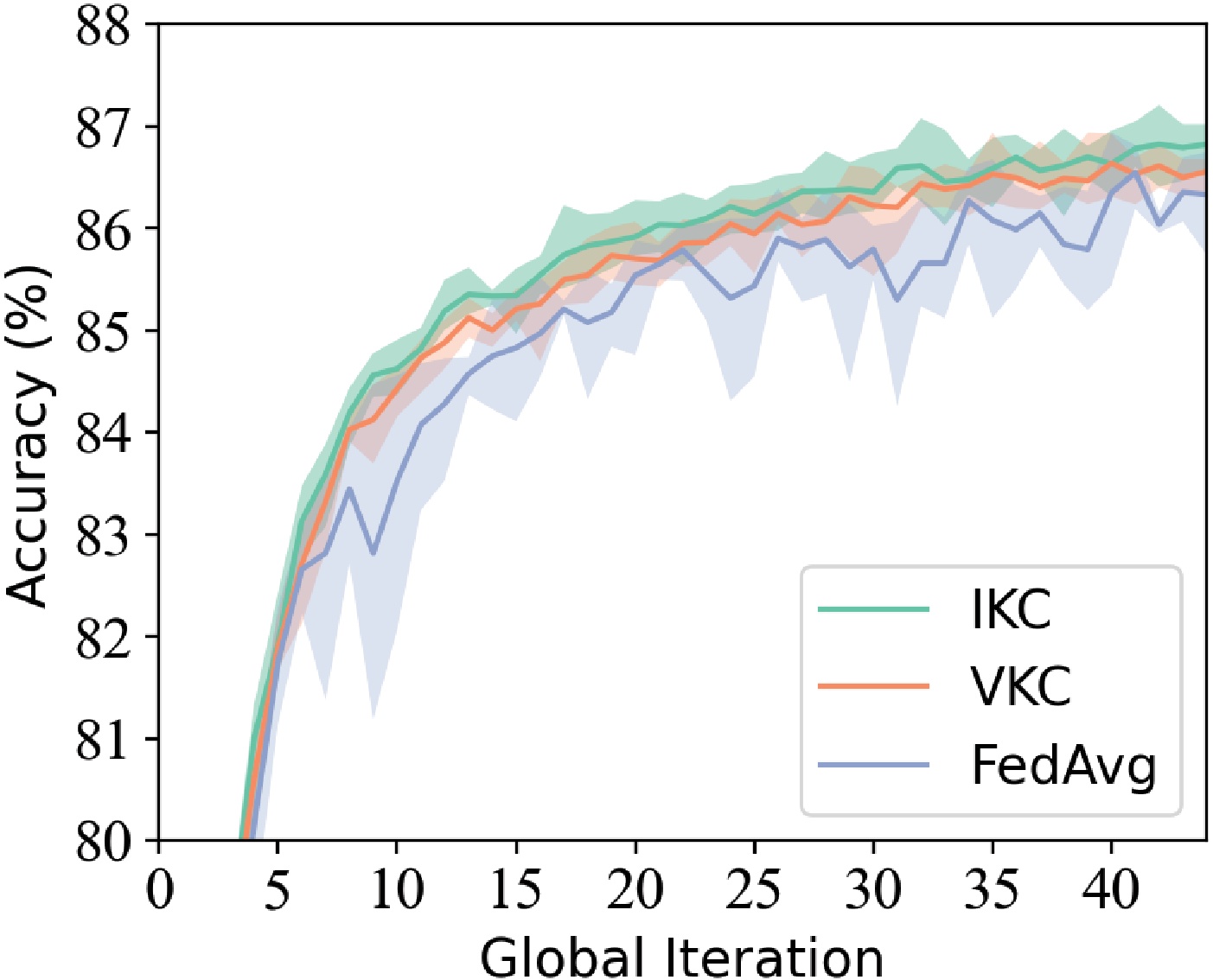}
%\caption{fig1}
\end{minipage}%
}\hspace{-0.5em}
\subfigure[$H = 30$]{
\begin{minipage}[t]{0.25\linewidth}
\centering
\includegraphics[width=1\linewidth]{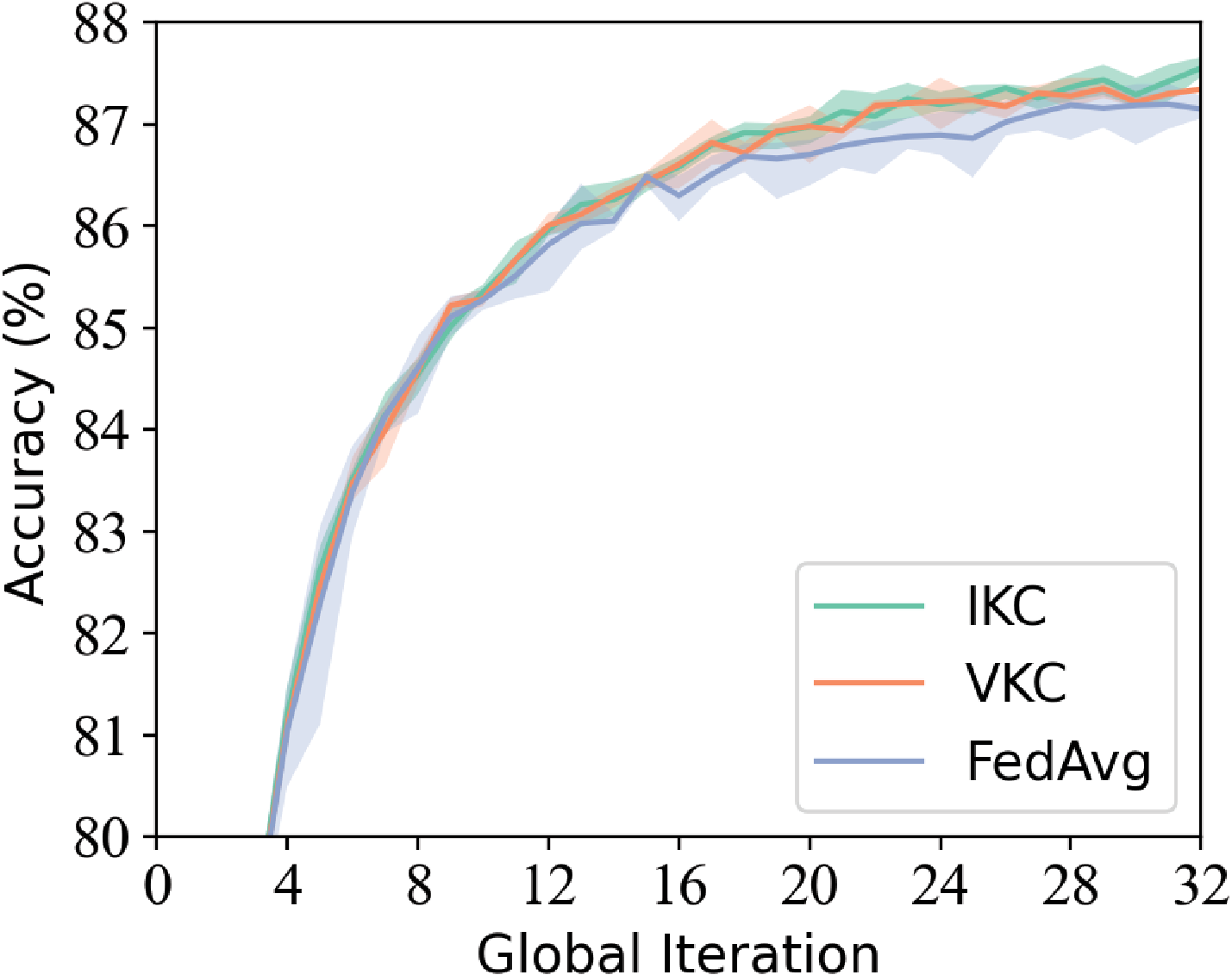}
%\caption{fig2}
\end{minipage}%
}\hspace{-0.8em}
\subfigure[$H = 50$]{
\begin{minipage}[t]{0.25\linewidth}
\centering
\includegraphics[width=1\linewidth]{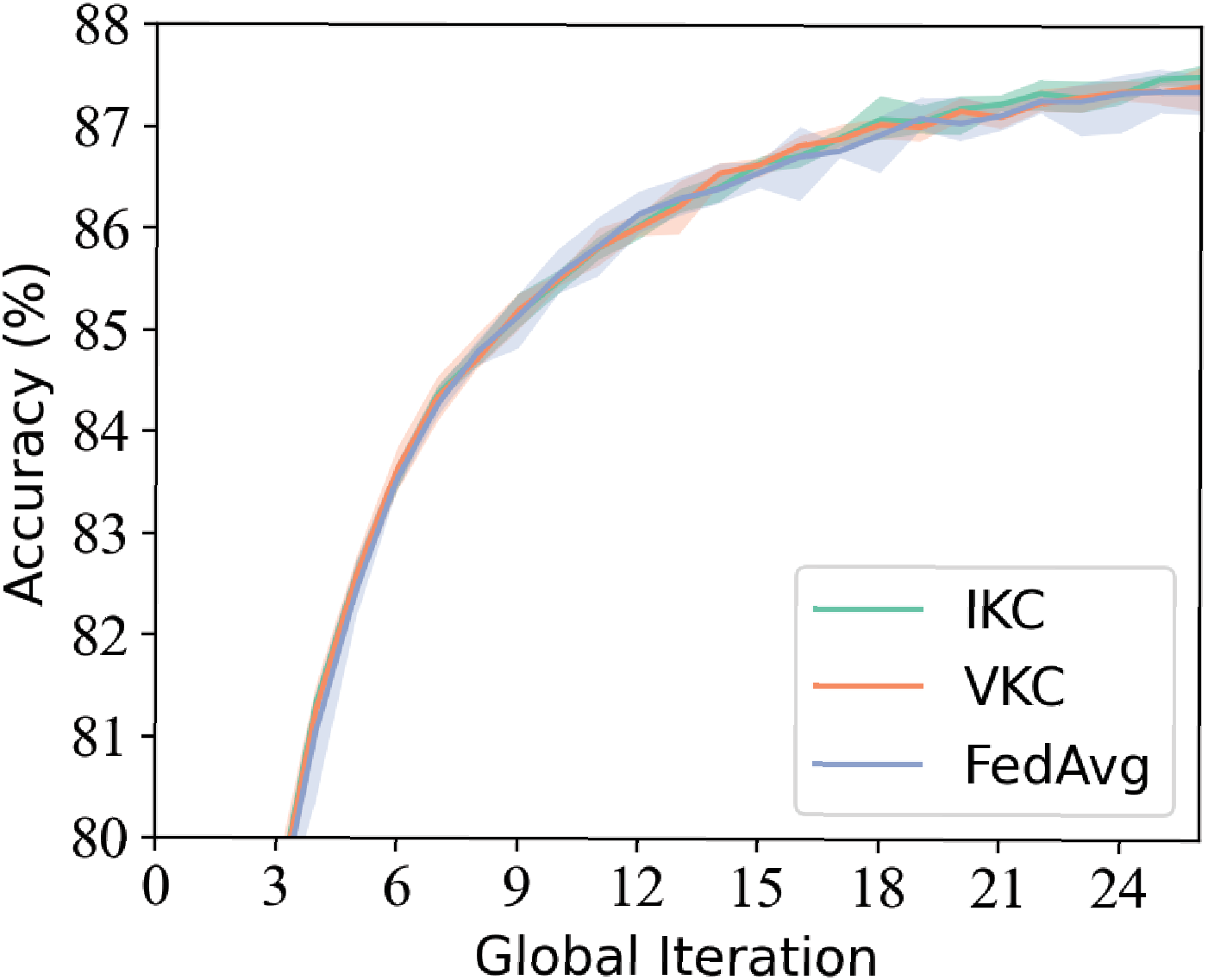}
%\caption{fig2}
\end{minipage}%
}\hspace{-0.5em}
\subfigure[$H = 70$]{
\begin{minipage}[t]{0.25\linewidth}
\centering
\includegraphics[width=1\linewidth]{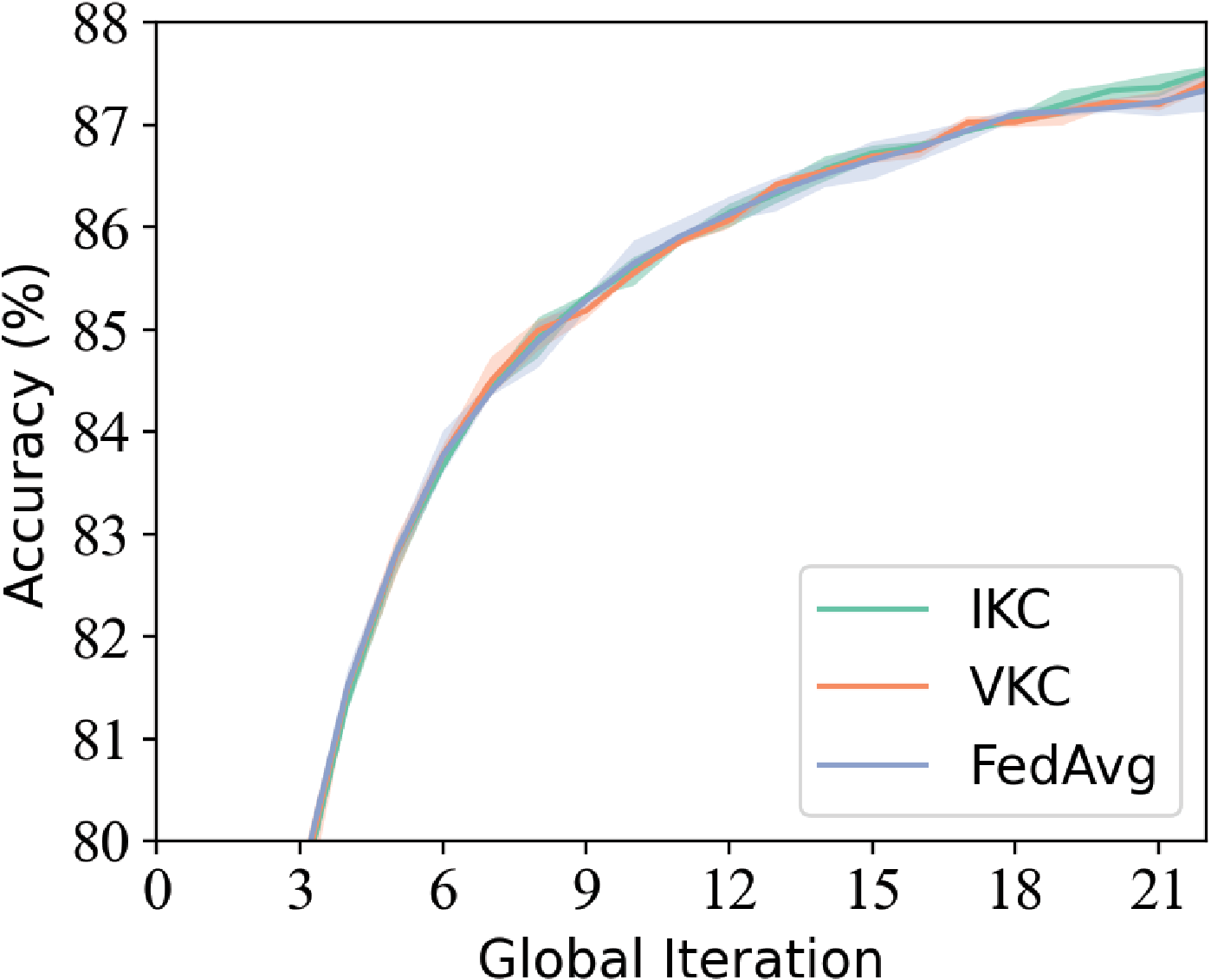}
%\caption{fig2}
\end{minipage}%
} \vspace{0pt}
\caption{Testing accuracy of HFL on FashionMNIST with different size of $\mathcal{H}$.}
\label{diff_S_fashion}
\vspace{0em}
\end{figure*}

\begin{figure*}[t]
\centering
\hspace{-1em}
\subfigure[$H = 10$]{
\begin{minipage}[t]{0.25\linewidth}
\centering
\includegraphics[width=1\linewidth]{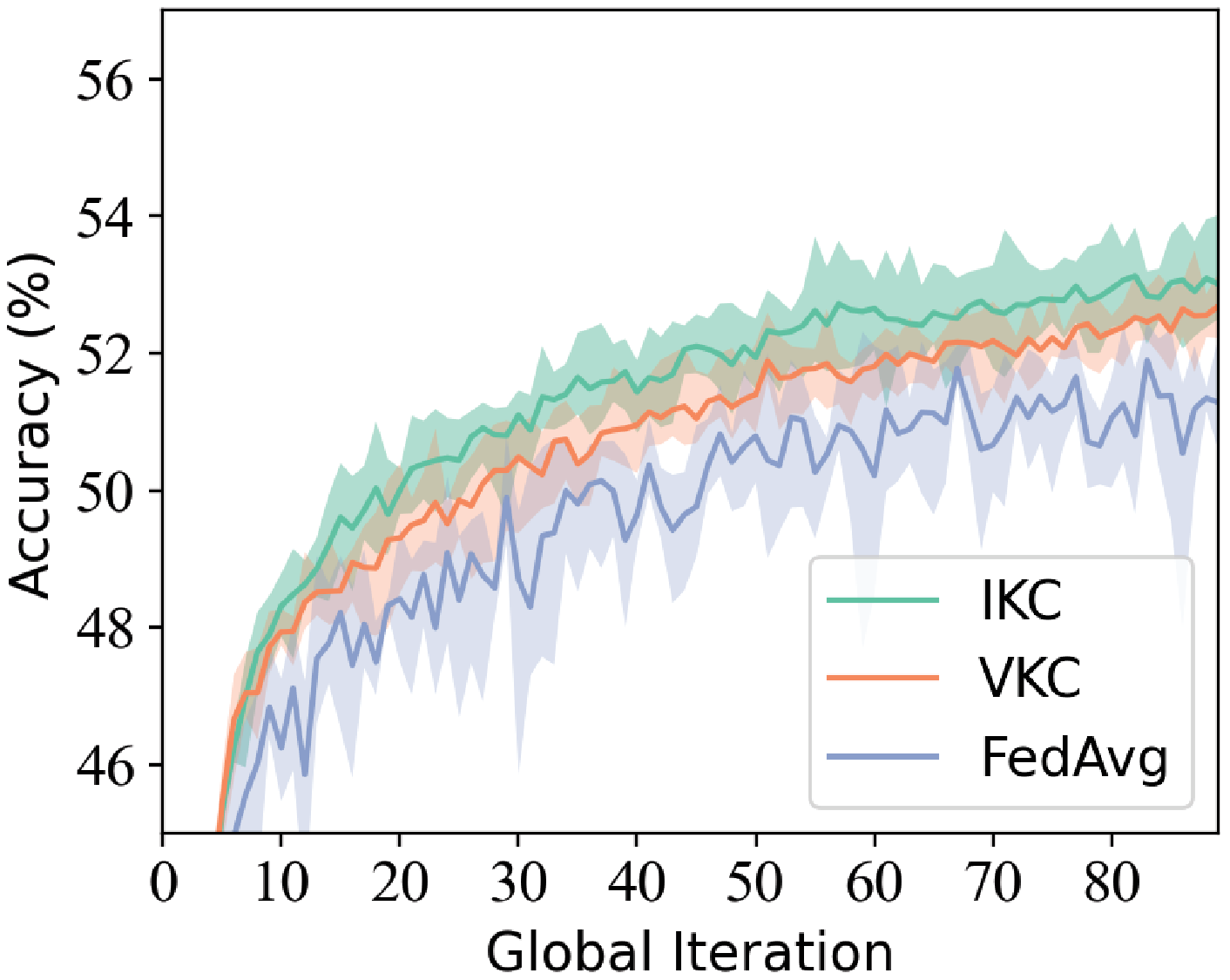}
%\caption{fig1}
\end{minipage}%
}\hspace{-0.5em}
\subfigure[$H = 30$]{
\begin{minipage}[t]{0.25\linewidth}
\centering
\includegraphics[width=1\linewidth]{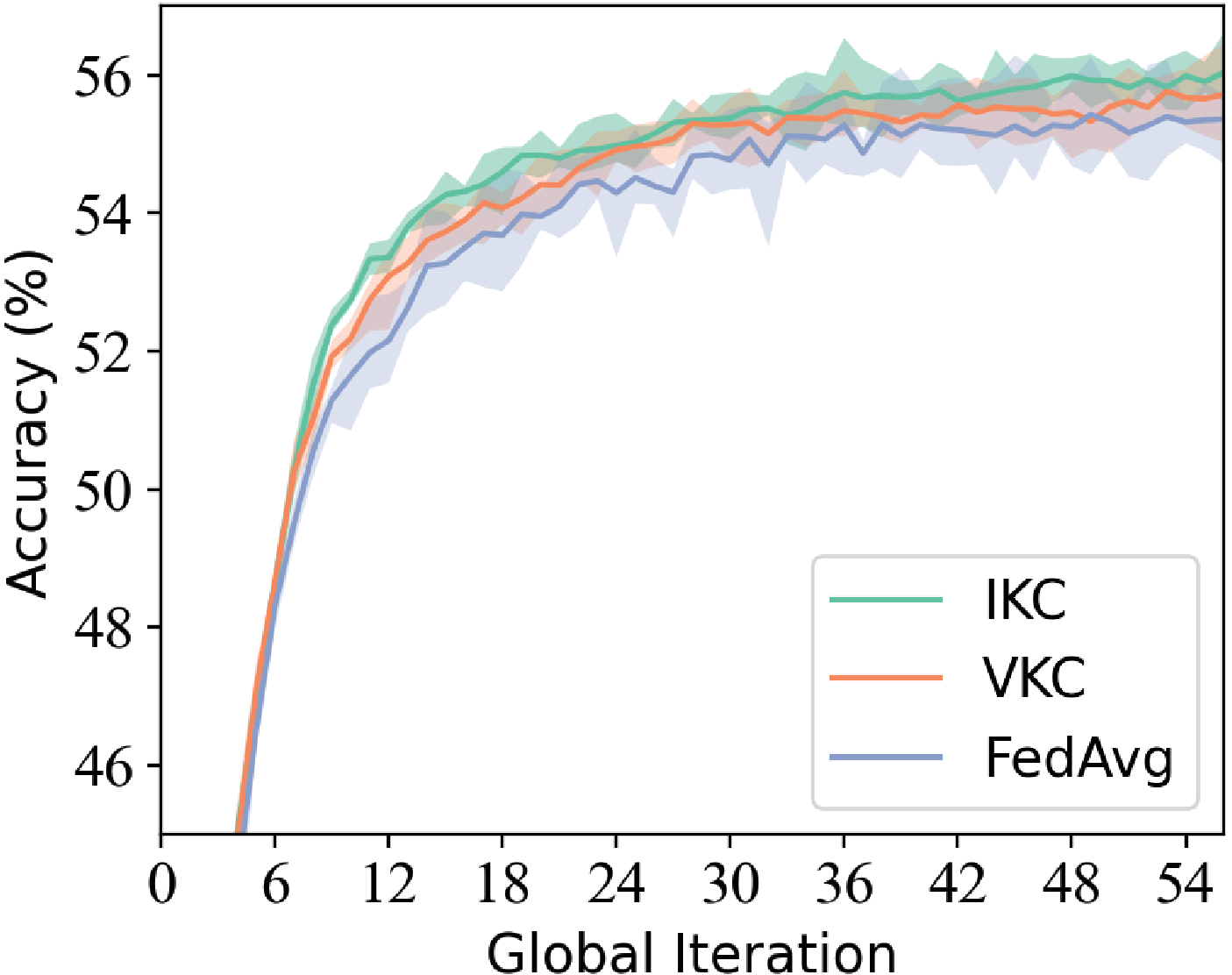}
%\caption{fig2}
\end{minipage}%
}\hspace{-0.8em}
\subfigure[$H = 50$]{
\begin{minipage}[t]{0.25\linewidth}
\centering
\includegraphics[width=1\linewidth]{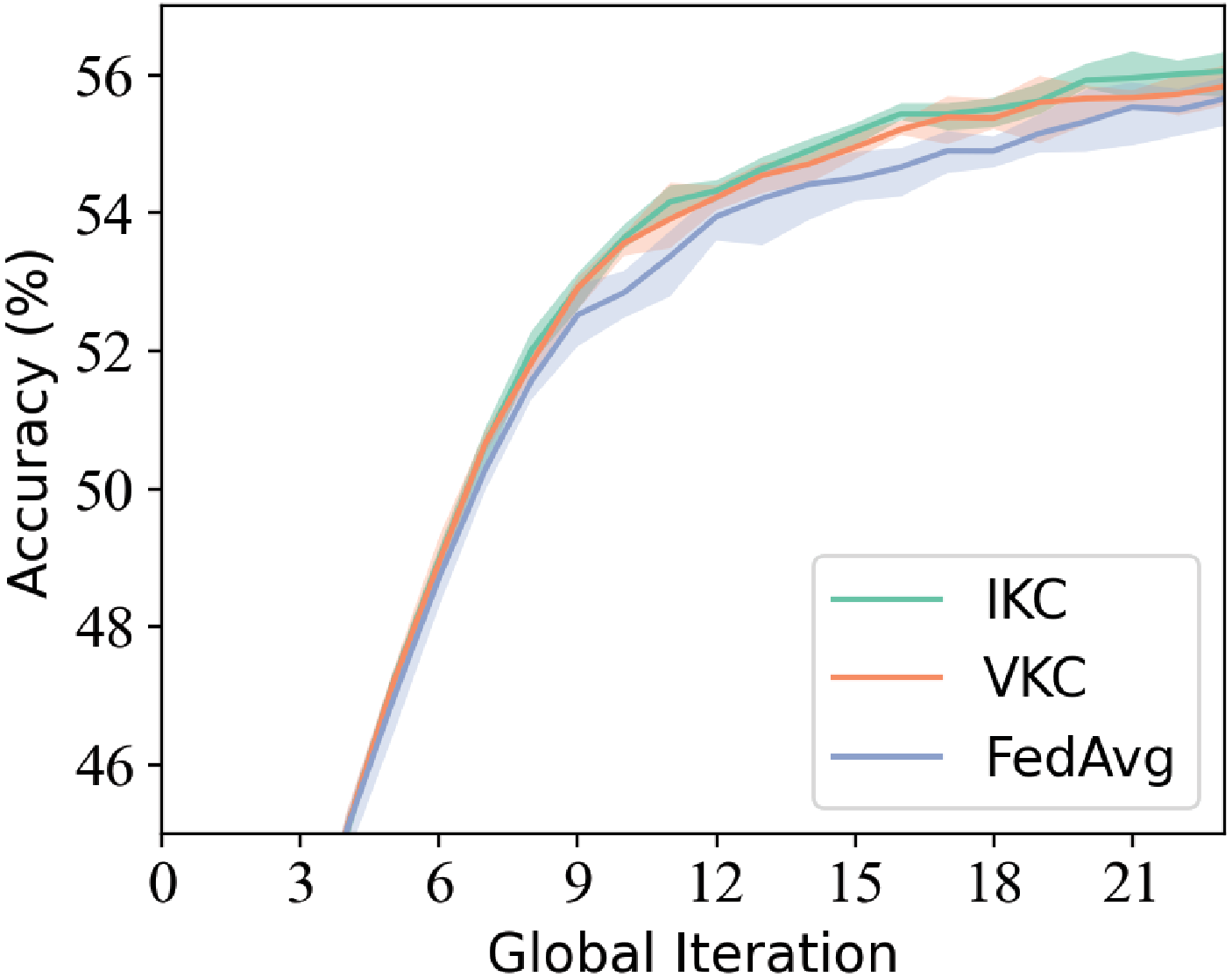}
%\caption{fig2}
\end{minipage}%
}\hspace{-0.5em}
\subfigure[$H = 70$]{
\begin{minipage}[t]{0.25\linewidth}
\centering
\includegraphics[width=1\linewidth]{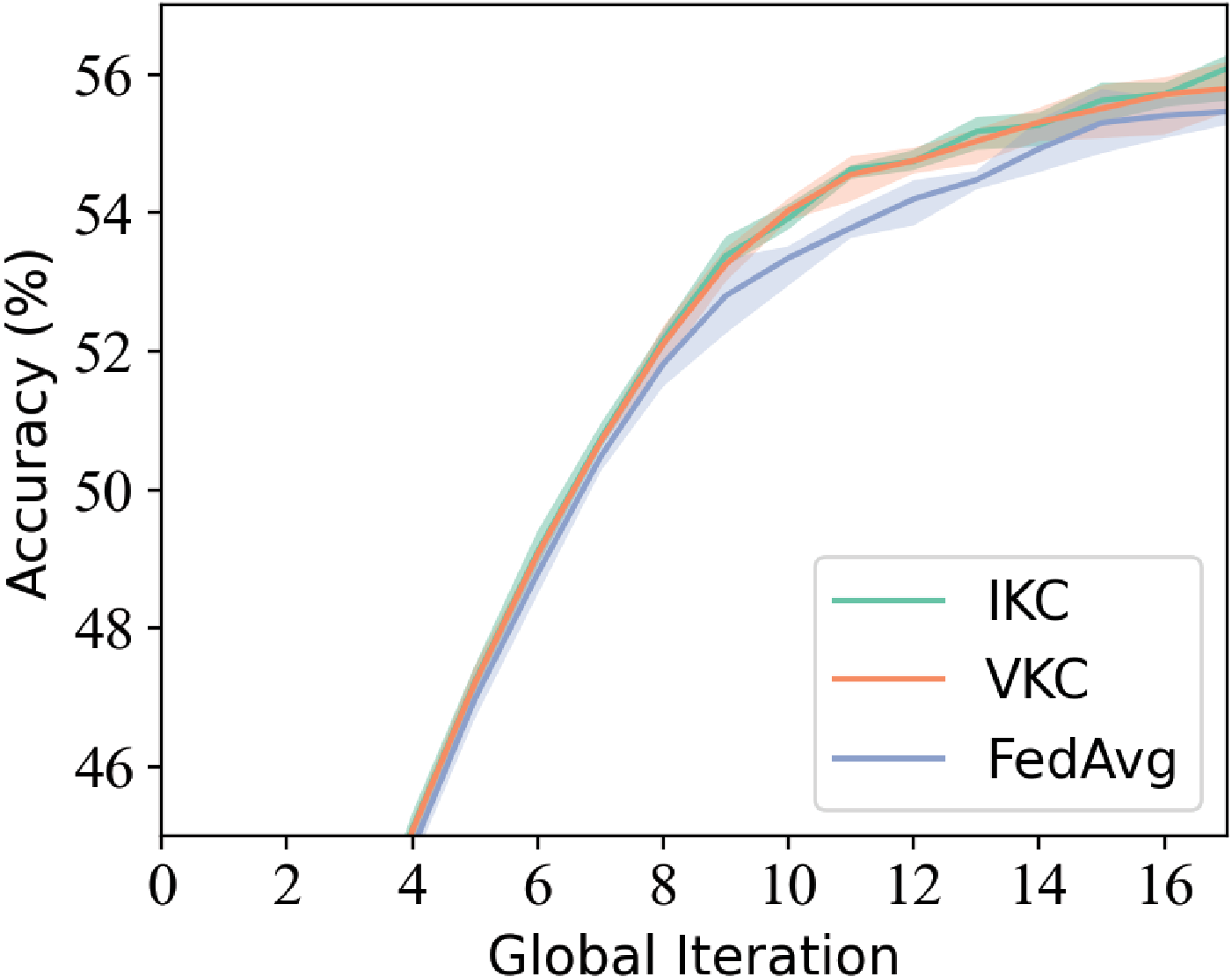}
%\caption{fig2}
\end{minipage}%
}%
\caption{Testing accuracy of HFL on CIFAR-10 with different size of $\mathcal{H}$.}
\label{diff_S_cifar}
\vspace{0em}
\end{figure*}

The performance of IKC is evaluated on FashionMNIST and CIFAR-10 using different sizes of $\mathcal{H}$. We compare the results of IKC with two benchmark models, FedAvg and VKC. To ensure the robustness of our findings, we train the HFL model five times, and the testing accuracies of the HFL model are presented in Fig.~\ref{diff_S_fashion} and Fig.~\ref{diff_S_cifar}, respectively. The solid curves depict the average accuracy, while the shaded area represents the standard deviation across the five experiments.

Holistically, increasing the number of scheduled devices at each iteration results in achieving equivalent testing accuracy with fewer global iterations according to Fig.~\ref{diff_S_fashion} and Fig.~\ref{diff_S_cifar}. Besides, scheduling more devices leads to a more stable and robust training process as the HFL model can acquire more knowledge from the devices, thereby improving the generalization capability of the model. More importantly, we see that the learning curves as well as the shaded areas of IKC are situated above those of VKC and FedAvg, which demonstrates that our method outperforms the two benchmarks in terms of accelerating the convergence of HFL. In addition, the performance difference among the three algorithms becomes smaller as $H$ grows. This is because a larger $H$ increases the possibility for IKC, VKC, and FedAvg to schedule the same devices.

We further compare the performance of IKC and VKC, and the evaluation criteria include Adjusted Rand index (ARI) which evaluates the clustering accuracy~\cite{Nguyen10} as well as time delay and energy consumption for training and transmitting the auxiliary model. ARI, which is defined in~\eqref{ARI_equa}, measures the similarity between the predicted cluster labels and the ground truth:
\begin{align}
ARI(\bm{\overline{\mathcal{C}}}, \bm{\widehat{\mathcal{C}}}) = &[2(\sigma_{00}\sigma_{11}-\sigma_{01}\sigma_{10})]/[(\sigma_{00}+\sigma_{01})(\sigma_{01}+\sigma_{11})\notag\\
&+(\sigma_{00}+\sigma_{10})(\sigma_{10}+\sigma_{11})],
\label{ARI_equa}
\end{align}
where $\bm{\overline{\mathcal{C}}} = \{\overline{\mathcal{C}}_1,...,\overline{\mathcal{C}}_c\}$ is the predicting clustering result, $\bm{\widehat{\mathcal{C}}} = \{\widehat{\mathcal{C}}_1,...,\widehat{\mathcal{C}}_c\}$ is the ground truth, $\sigma_{11}$ is the number of pairs that are in the same cluster in both $\bm{\overline{\mathcal{C}}}$ and $\bm{\widehat{\mathcal{C}}}$, $\sigma_{00}$ is the number of pairs that are in different clusters in both $\bm{\overline{\mathcal{C}}}$ and $\bm{\widehat{\mathcal{C}}}$, $\sigma_{01}$ is the number of pairs that are in the same cluster in $\bm{\overline{\mathcal{C}}}$ but in different clusters in $\bm{\widehat{\mathcal{C}}}$, and $\sigma_{10}$ is the number of pairs that are in different clusters in $\bm{\overline{\mathcal{C}}}$ but in the same cluster in $\bm{\widehat{\mathcal{C}}}$. A high ARI value indicates a strong agreement between the predicted clustering and the ground truth, with a maximum value of 1 indicating identical clusters and a value close to 0 indicating little to no agreement. Table~\ref{clustering} depicts the time delay and energy consumption for performing Algorithm~\ref{KMeans} as well as the ARI values. It can be noted that IKC completes the clustering process using the lowest time delay and energy consumption. This is because the mini model $\bm{\xi}$ is much smaller than the original HFL model, thereby alleviating the communication and computation overheads. Besides, the ARI values of both IKC and VKC are 1. This indicates that the mini model $\bm{\xi}$ is sufficient for Algorithm~\ref{KMeans} to correctly cluster the IoT devices based on their majority classes.

\vspace{0em}

\begin{table}[t]
\centering
\caption{Time delay and energy consumption for conducting Algorithm~\ref{KMeans} and the ARI values}
\textsl{}\setstretch{1}
\setlength{\tabcolsep}{1.2mm}{
\begin{tabular}{|c|c|c|c|}
\hline
\diagbox{Methods}{\makecell[c]{Criteria}} & \begin{tabular}[c]{@{}c@{}}Time delay \\ (s)\end{tabular} & \begin{tabular}[c]{@{}c@{}}Energy \\ consumption (J)\end{tabular} & ARI \\ \hline
IKC                                                          & 3.1                                                       & 23.5                                                              & 1.0 \\ \hline
\begin{tabular}[c]{@{}c@{}}VKC\\ (FashionMNIST)\end{tabular} & 128.0                                                     & 671.0                                                           & 1.0 \\ \hline
\begin{tabular}[c]{@{}c@{}}VKC\\ (CIFAR-10)\end{tabular}     & 252.6                                                     & 1317.0                                                             & 1.0 \\ \hline
\end{tabular}}
\label{clustering}
\vspace{0em}
\end{table}

\begin{figure}[t]
  \centering
  \includegraphics[width=1\linewidth]{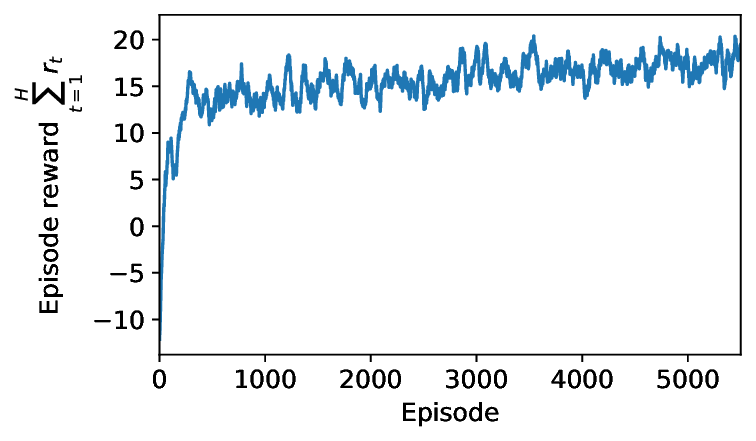}\vspace{-10pt}
\caption{Learning curve of the proposed D$^3$QN.}
\label{lstm}
\vspace{0em}
\end{figure}

\begin{figure*}[!t]
\hspace{-2em}
\subfigure[Time delay $T_i$]{
\begin{minipage}[t]{0.3\linewidth}
\centering
\includegraphics[width=5.7cm]{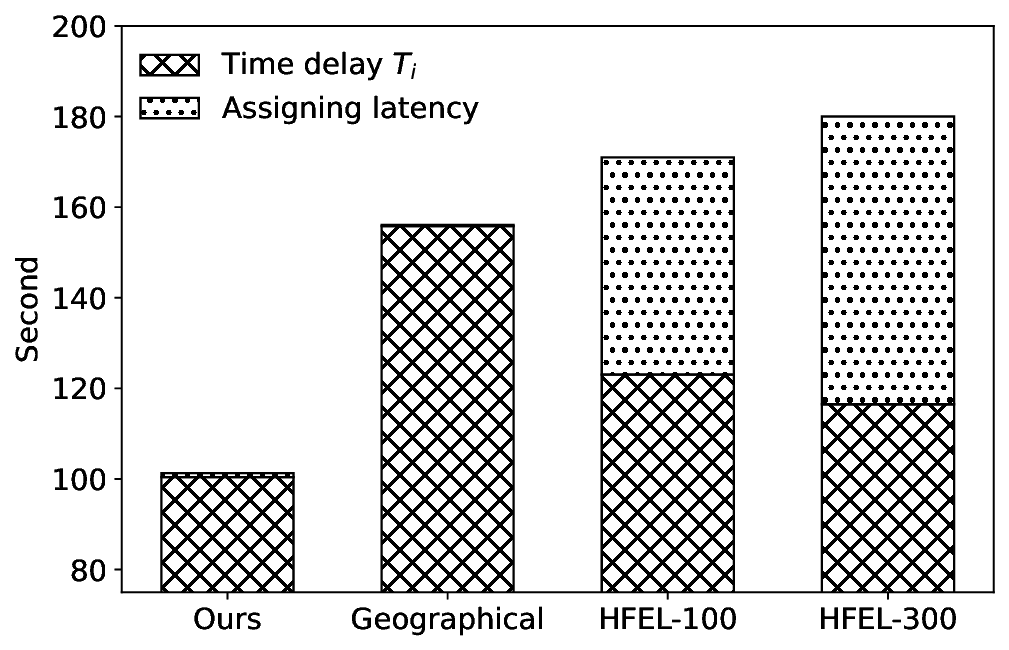}
%\caption{fig1}
\end{minipage}%
\label{oneET_1}
}\hspace{0.8em}
\subfigure[Energy consumption $E_i$]{
\begin{minipage}[t]{0.3\linewidth}
\centering
\includegraphics[width=5.7cm]{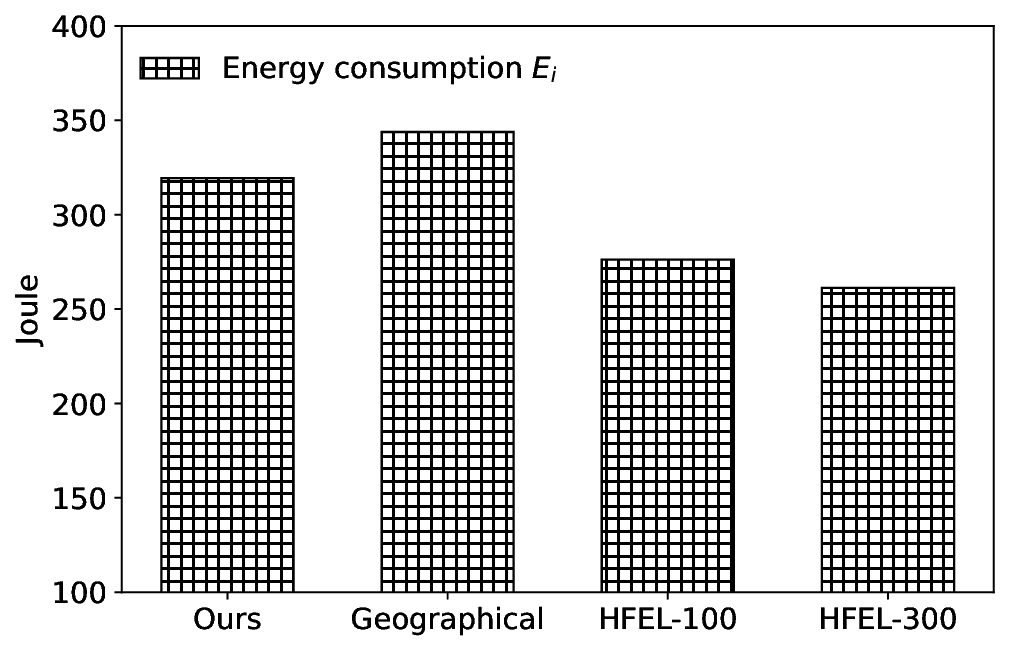}
%\caption{fig2}
\end{minipage}%
\label{oneET_2}
}\hspace{.3em}
\subfigure[Objective value $E_i$ (Joule) + $\lambda T_i$ (Second)]{
\begin{minipage}[t]{0.38\linewidth}
\centering
\includegraphics[width=5.7cm]{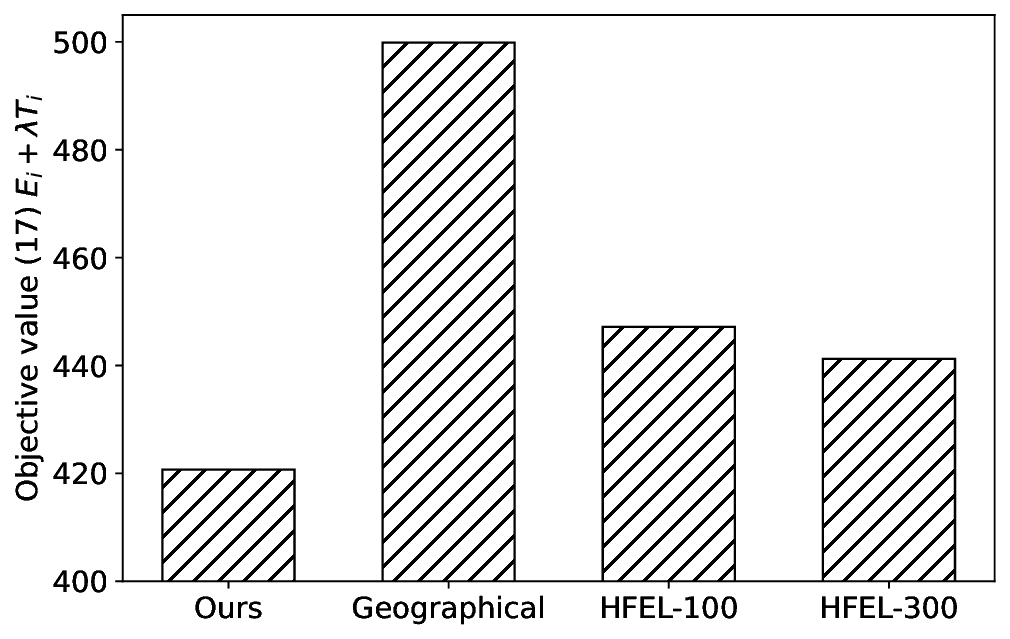}
%\caption{fig2}
\end{minipage}%
\label{oneET_3}
}\vspace{0pt}
\caption{Assignment results with different assignment strategies ($\lambda = 1$).}
\label{oneET}
\vspace{0em}
\end{figure*}

\begin{figure*}[!t]
\hspace{-2em}
\subfigure[FashionMNIST]{
\begin{minipage}[t]{0.3\linewidth}
\centering
\includegraphics[width=5.2cm]{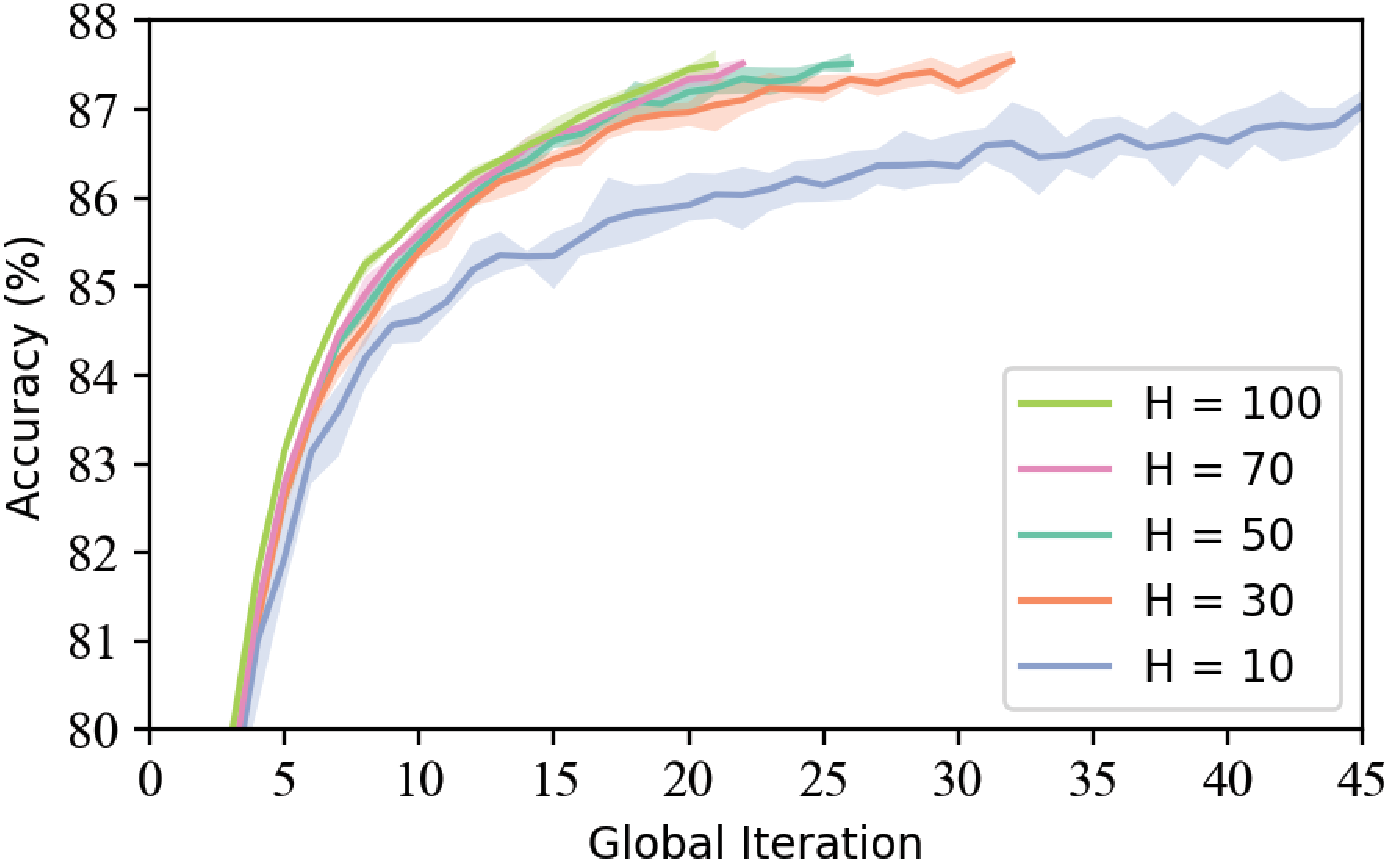}
%\caption{fig1}
\end{minipage}%
\label{overET_1}
}\hspace{0em}
\subfigure[CIFAR-10]{
\begin{minipage}[t]{0.3\linewidth}
\centering
\includegraphics[width=5.4cm]{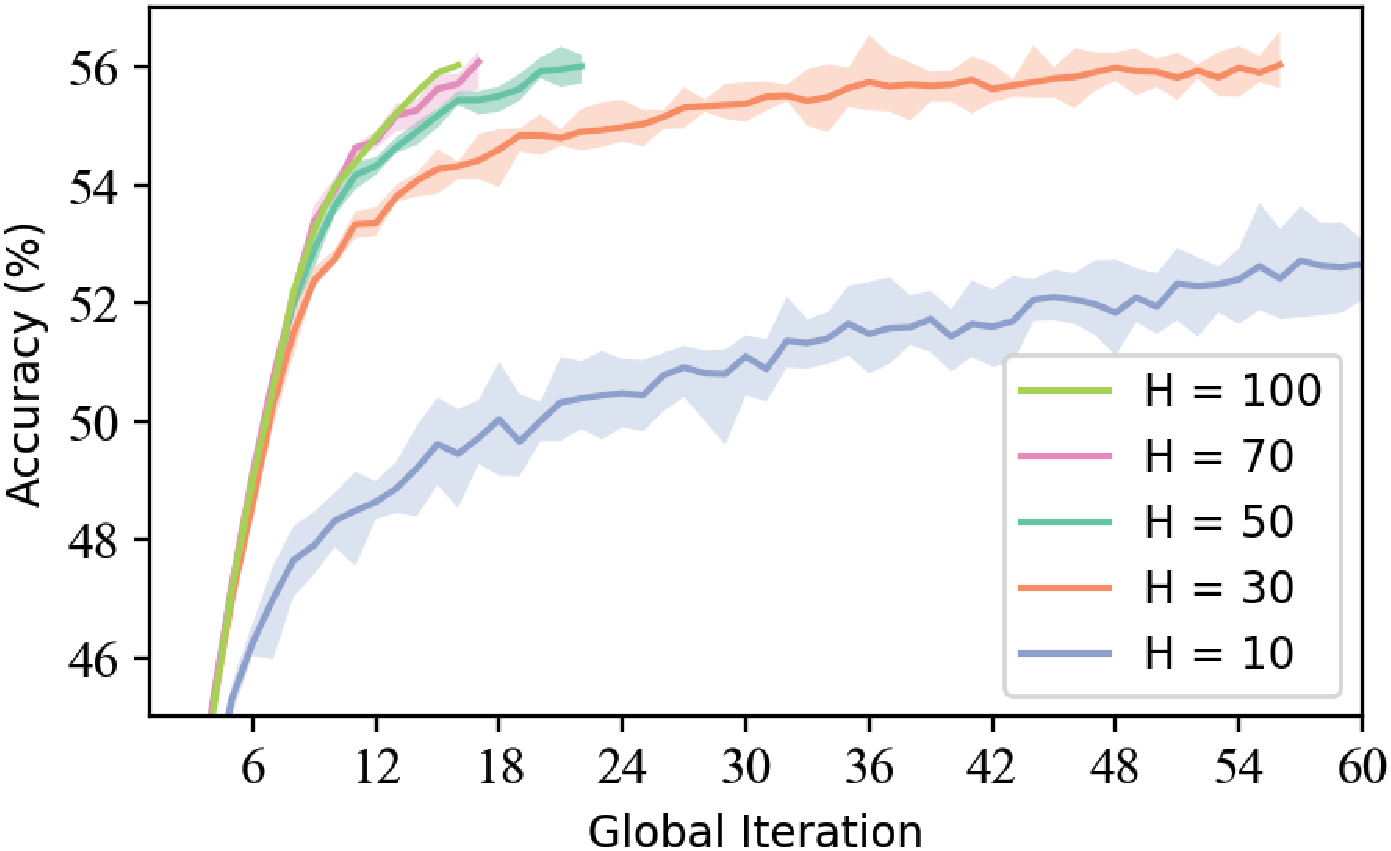}
%\caption{fig2}
\end{minipage}%
\label{overET_2}
}
\subfigure[Objective value $E$ $(\text{Joule})$ + $\gamma T$ $(\text{Second})$ ]{
\begin{minipage}[t]{0.35\linewidth}
\centering
\includegraphics[width=6.5cm]{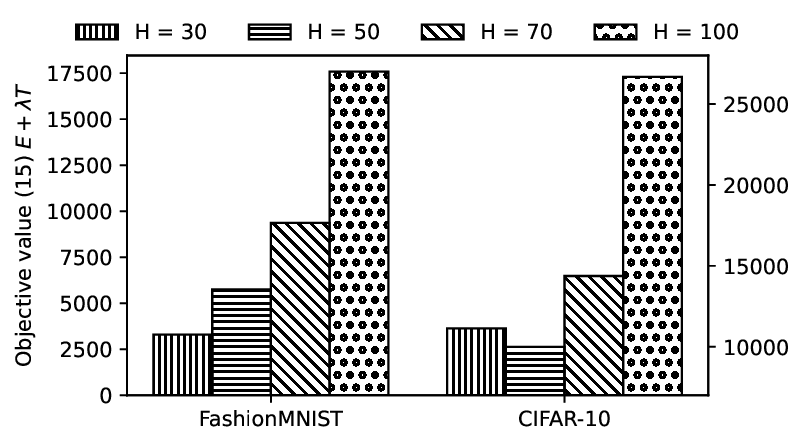}
%\caption{fig2}
\end{minipage}%
\label{overET_3}
}

\hspace{-1em}
\subfigure[Time delay $T$]{
\begin{minipage}[t]{0.23\linewidth}
\raggedright
\includegraphics[width=3.8cm]{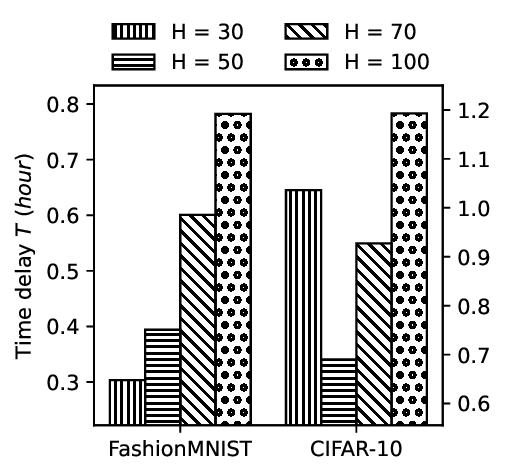}
%\caption{fig1}
\end{minipage}%
\label{overET_4}
}\hfil
\subfigure[Energy consumption $E$]{
\begin{minipage}[t]{0.23\linewidth}
\raggedright
\includegraphics[width=3.8cm]{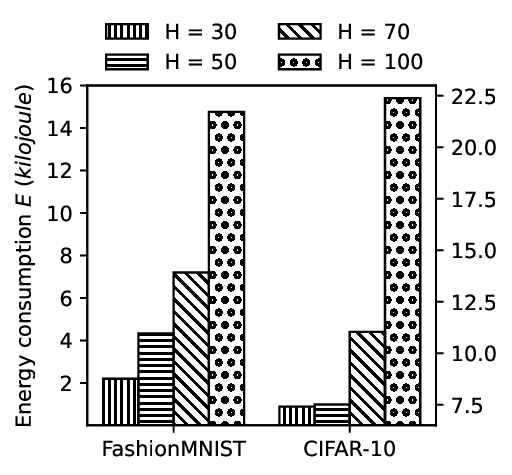}
%\caption{fig2}
\end{minipage}%
\label{overET_5}
}\hfil
\subfigure[Size of messages tra-\newline nsmitted per global iteration]{
\begin{minipage}[t]{0.23\linewidth}
\raggedright
\includegraphics[width=3.8cm]{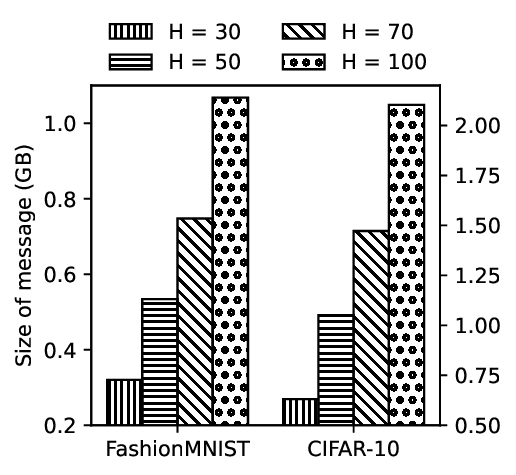}
\end{minipage}%
\label{overET_6}
}\hfil
\subfigure[Size of messages transmitted during the entire HFL algorithm]{
\begin{minipage}[t]{0.25\linewidth}
\raggedright
\includegraphics[width=3.8cm]{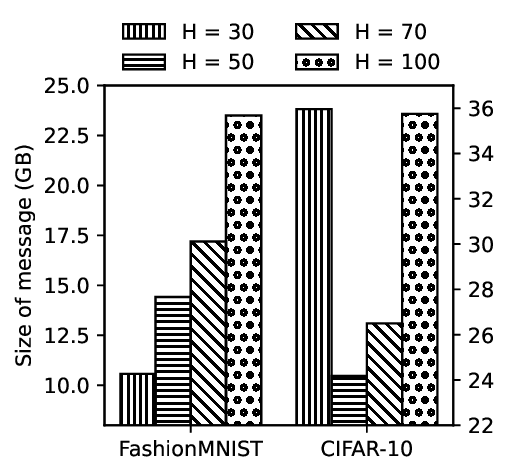}
%\caption{fig2}
\end{minipage}%
\label{overET_7}
} \vspace{0pt}
\caption{Testing accuracy of HFL, objective value~\eqref{YY}, time delay $T$, energy consumption $E$, and size of transmitted messages on FashionMNIST and CIFAR-10 with different $H$.}
\label{overET}
\vspace{0em}
\end{figure*}

\subsection{Evaluation of device assignment\\[0em]}\label{assign_result}

We train the DRL model based on Algorithm~\ref{TrainingDRL} with $H = 50$ and $\lambda = 1$. The values of the randomly-generated parameters $\big\{(\bar{g}^1_{n_t}, ..., \bar{g}^M_{n_t}, u_{n_t}, D_{n_t}, p_{n_t} )\big| t = 1,...,H \big\}$ at Line 4 of Algorithm~\ref{TrainingDRL} are constrained by Table~\ref{para_setup}. As the benchmark strategy, HFEL performs device transferring adjustment 100 times and device exchanging adjustment 300 times. 

Figure~\ref{lstm} illustrates the learning curves of the deep reinforcement learning (DRL) model. The y-axis of the figure displays the average accumulated reward over 50 episodes. As shown in the plot, the reward progressively increases from the beginning of the training process and eventually stabilizes around 17 when the algorithm converges. The experimental results have demonstrated that the proposed DRL algorithm can acquire the ability to make assignment decisions comparable to the HFEL strategy.

We evaluate the performance of the well-trained DRL model by comparing it with three benchmarks, namely HFEL-100, HFEL-300, and geographical distribution-based strategy. HFEL-100 and HFEL-300 refer to the HFEL strategies that perform device exchanging adjustments 100 and 300 times, respectively. Note that both HFEL-100 and HFEL-300 carry out device transferring adjustments 100 times. The geographical distribution-based strategy assigns devices to the edge server closest to them. We randomly generate $\big\{(\bar{g}^1_{n_t}, ..., \bar{g}^M_{n_t}, u_{n_t}, D_{n_t}, p_{n_t} )\big| t = 1,...,H \big\}$ for 100 iterations, during which we employ various strategies to conduct device assignments for each iteration. 

Figure~\ref{oneET} provides the average experimental results, including time delay $T_i$, energy consumption $E_i$, and the objective value $E_i + \lambda T_i$. According to Fig.~\ref{oneET_1} and Fig.~\ref{oneET_2}, our proposed method achieves the lowest time delay $T_i$, while HFEL-300 displays the lowest energy consumption. Furthermore, both our proposed method and the geographical distribution-based strategy display small assigning latency, whereas the HFEL strategy suffers from high assigning latency. As shown in Figure~\ref{oneET_3}, our proposed method achieves the lowest objective value in comparison to other benchmarks. In summary, the proposed method attains comparable performance to HFEL-300 while necessitating less latency for executing the assigning process.

\vspace{0em}

\subsection{Evaluation of the proposed HFL framework\\[0em]}\label{interplay}

The proposed HFL framework is evaluated using Algorithm~\ref{Overall_framework} with varying sizes of $\mathcal{H}$. Note that scheduling all the devices (i.e., $H = 100$) can be regarded as the traditional HFL method. We introduce target accuracies as the criterion for justifying the convergence of HFL. The target accuracies are set to 87.5\% and 56\% for FashionMNIST and CIFAR-10, respectively. The HFL framework is trained five times for each $H$. Testing accuracy, objective value~\eqref{YY}, time delay $T$, energy consumption $E$, and size of transmitted messages are shown in Fig.~\ref{overET}.

According to Fig.~\ref{overET_1} and~\ref{overET_2}, device scheduling requires HFL to use more global iterations for reaching the target accuracy. As shown in Fig.~\ref{overET_3}-\ref{overET_5}, however, the completion time $T$ and the total energy consumption $E$ increase significantly if all devices participate in the local update, which leads to the highest objective value~\eqref{YY}. The results indicate that the proposed HFL framework effectively decreases the objective value~\eqref{YY}, thus confirming the necessity of device scheduling in HFL.

In addition, the value of $H$ significantly affects the performance of HFL and thus should be carefully chosen. Basically, the optimal value of $H$ depends on two factors: the target accuracy and the complexity of the datasets. While a larger value of $H$ may lead to increased communication and computation overheads, a small $H$ may be insufficient to achieve the target accuracy.  For example, in Fig.~\ref{overET_1} and~\ref{overET_2}, $H = 10$ is inadequate for achieving the target accuracy. Regarding the complexity of the dataset, it can be observed that FashionMNIST achieves the minimal objective value under $H=30$, while the optimal $H$ for CIFAR-10 is 50. This is because the HFL model trained on CIFAR-10 requires much more global iterations to achieve the target accuracy when $H=30$ than when $H=50$. In contrast, FashionMNIST is less complex than CIFAR-10, and an $H$ value of 30 is sufficient to achieve the target accuracy within a similar number of global iterations to that obtained with $H=50$. 

Fig.~\ref{overET_6} and~\ref{overET_7} depict the size of messages that are transmitted per global iteration and through the entire training process, respectively. It is evident that reducing the number of scheduled devices significantly decreases the size of messages transmitted per global iteration, consequently easing network congestion. Nonetheless, when training with CIFAR-10 with $H=30$, the largest size of messages transmitted through the entire HFL algorithm is obtained due to more global iterations. Based on the experimental findings, it can be deduced that scheduling 50\% of the devices is a suitable approach for general cases. However, if the practical scenario demands specific criteria for time delay or the size of messages transmitted per global round, then scheduling 30\% of the devices is the recommended approach.

% \begin{figure}[t]
%   \centering
%   \includegraphics[width=\linewidth]{LSTM.pdf}
% \caption{Workflow of the BiLSTMs-based model at the $t$-th time slot. $\phi$ represents the parameters of the LSTM modules.}
% \label{lstm}
% \end{figure}

\vspace{0em}

\section{Conclusions\\[0em]}\label{section8}

In this paper, we have studied the mechanism of HFL and investigated device scheduling and assignment in HFL. We formulate a joint optimization problem to minimize the weights sum of the time delay and energy consumption for training the entire HFL in the IoT system. We propose a novel HFL framework to solve this problem effectively. The proposed framework employs the improved K-center (IKC) algorithm for device scheduling and the dueling double deep Q-Network (D$^3$QN) for device assignment. We compare the proposed methods with several baselines and evaluate the proposed HFL framework on FashionMNIST and CIFAR-10. The experimental results show that the proposed IKC algorithm enables HFL to achieve the target accuracy within fewer global iterations, while the DRL method can conduct fast device assignment and allows HFL to obtain low time delay and energy consumption. The whole HFL framework effectively reduces the time delay and energy consumption required for training the entire HFL algorithm when compared with traditional HFL. Additionally, scheduling 50\% of the IoT devices is generally sufficient to achieve convergence in HFL while decreasing the time delay and energy consumption. If reducing energy consumption and the size of transmitted messages in each round is a primary concern, it is recommended to schedule only 30\% of the devices.

% \section*{Acknowledgements}

% This research is supported by the National Research Foundation, Singapore and Infocomm Media Development Authority under its Trust Tech Funding Initiative and Strategic Capability Research Centres Funding Initiative. Any opinions, findings and conclusions or recommendations expressed in this material are those of the author(s) and do not reflect the views of National Research Foundation, Singapore and Infocomm Media Development Authority.

\linespread{1.0}

%\bibliographystyle{IEEEtran}
%\bibliography{sample-base}

\vspace{-1em}
\begin{IEEEbiography}[{\includegraphics[width=1in,clip,keepaspectratio]{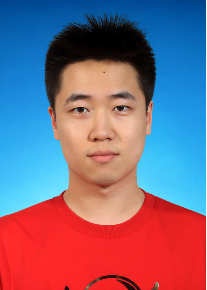}}]{Tinghao Zhang} is currently a Ph.D student in School of Computer Science and Engineering at Nanyang Technological University, Singapore. His main research interests include federated learning, edge computing, and wireless communication. He obtained his B.S. degree in the School of Electrical Engineering and Automation and M.S. degree in the School of Instrumentation Science and Engineering from Harbin Institute of Technology in 2014 and 2018, respectively.
\end{IEEEbiography}
\begin{IEEEbiography}[{\includegraphics[width=1in,clip,keepaspectratio]{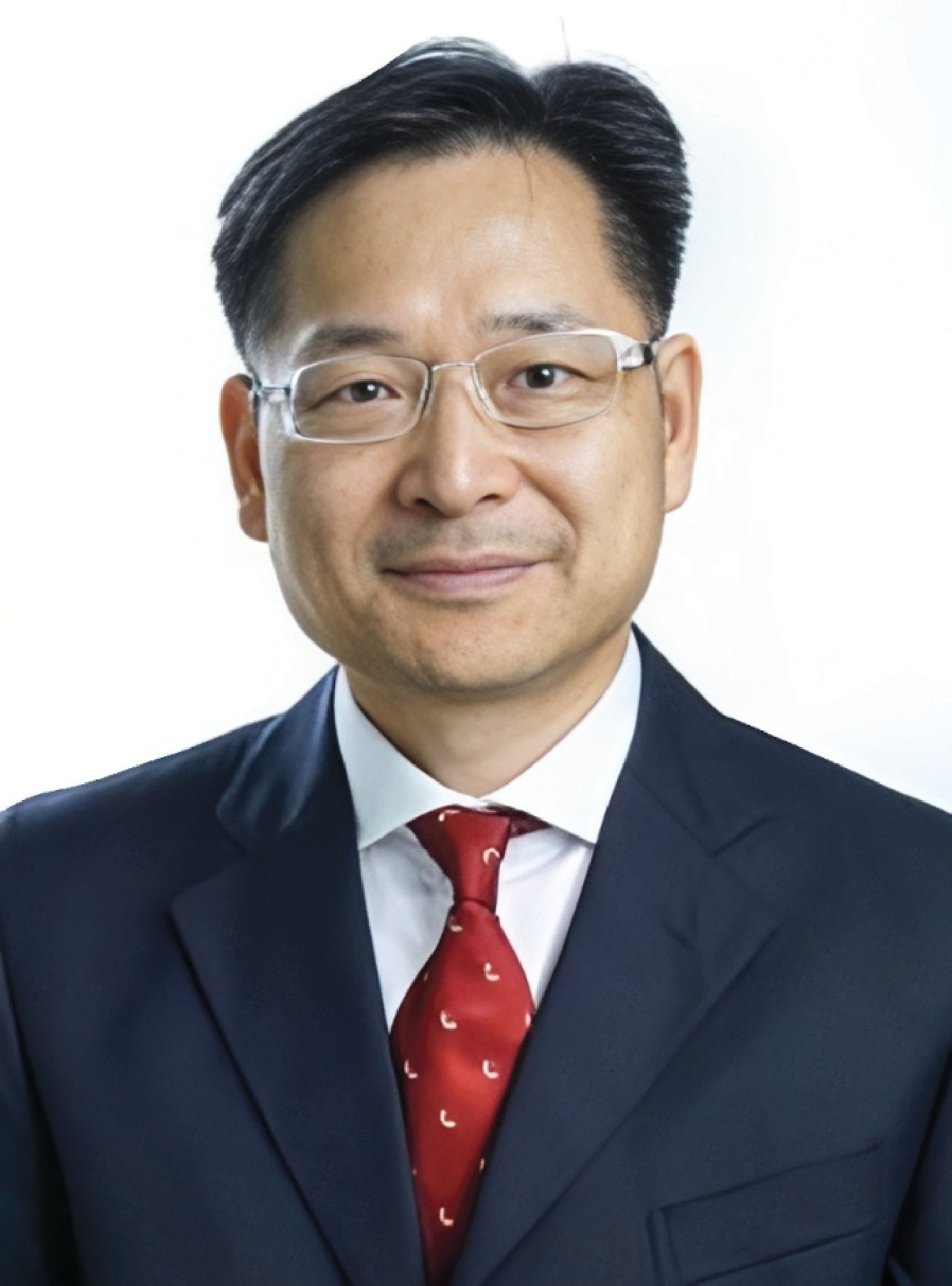}}]{Kwok-Yan Lam}(Senior Member, IEEE) received his B.Sc. degree ($1^\text{st}$ Class Hons.) from University of London, in 1987, and Ph.D. degree from University of Cambridge, in 1990. He is the Associate Vice President (Strategy and Partnerships) and Professor in the School of Computer Science and Engineering at the Nanyang Technological University, Singapore. He is currently also the Executive Director of the National Centre for Research in Digital Trust, and Director of the Strategic Centre for Research in Privacy-Preserving Technologies and Systems (SCRiPTS). From August 2020, he is on part-time secondment to the INTERPOL as a Consultant at Cyber and New Technology Innovation. Prior to joining NTU, he has been a Professor of the Tsinghua University, PR China (2002–2010) and a faculty member of the National University of Singapore and the University of London since 1990. He was a Visiting Scientist at the Isaac Newton Institute, Cambridge University, and a Visiting Professor at the European Institute for Systems Security. In 1998, he received the Singapore Foundation Award from the Japanese Chamber of Commerce and Industry in recognition of his research and development achievement in information security in Singapore. He is the recipient of the Singapore Cybersecurity Hall of Fame Award in 2022. His research interests include Distributed Systems, Intelligent Systems, IoT Security, Distributed Protocols for Blockchain, Homeland Security and Cybersecurity.
\end{IEEEbiography}
\begin{IEEEbiography}[{\includegraphics[width=1in,clip,keepaspectratio]{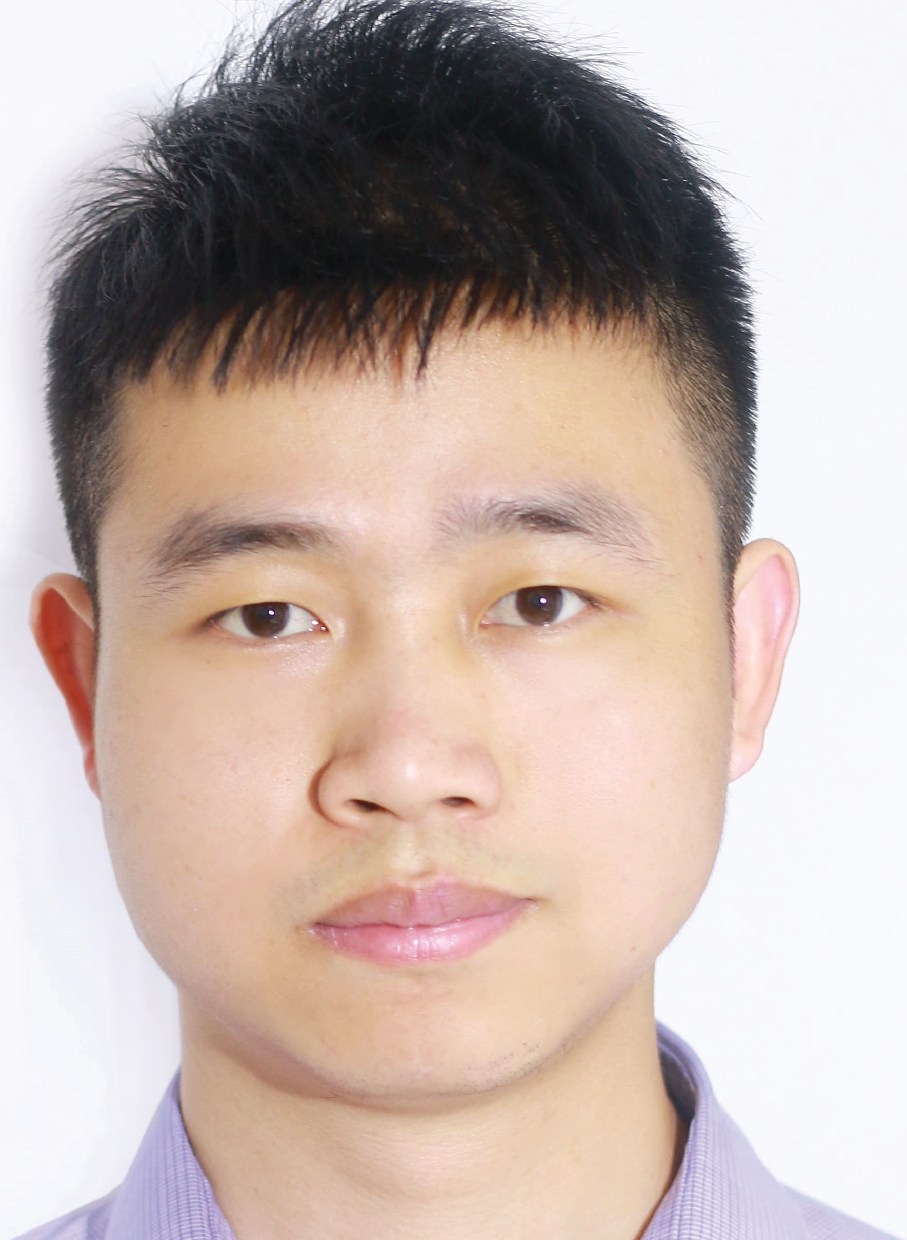}}]{Jun Zhao}(Senior Member, IEEE) (S'10-M'15) is currently an Assistant Professor in the School of Computer Science and Engineering (SCSE) at Nanyang Technological University (NTU) in Singapore. He received a PhD degree in May 2015 in Electrical and Computer Engineering from Carnegie Mellon University (CMU) in the USA (advisors: Virgil Gligor, Osman Yagan; collaborator: Adrian Perrig), affiliating with CMU's renowned CyLab Security \& Privacy Institute, and a bachelor's degree in July 2010 from Shanghai Jiao Tong University in China. Before joining NTU first as a postdoc with Xiaokui Xiao and then as a faculty member, he was a postdoc at Arizona State University as an Arizona Computing PostDoc Best Practices Fellow (advisors: Junshan Zhang, Vincent Poor).
\end{IEEEbiography}

% Generated by IEEEtran.bst, version: 1.14 (2015/08/26)

\end{document}